\documentclass[a4paper,fleqn,usenatbib]{mnras}
\usepackage{newtxtext,newtxmath}
\usepackage[T1]{fontenc}
\usepackage{ae,aecompl}
\usepackage{booktabs}	
\usepackage{graphicx}	
\usepackage{amsmath}	
\usepackage{amssymb}	
\usepackage{aas_macros} 
\usepackage[usenames]{color}

\usepackage{tablefootnote,footnotehyper} 
\usepackage{array}
\newcolumntype{H}{>{\setbox0=\hbox\bgroup}c<{\egroup}@{}} 

\title[The $\delta$\,Sct--roAp hybrid KIC\,11296437]{On the first $\delta$\,Sct--roAp hybrid pulsator and the stability of p and g modes in chemically peculiar A/F stars}
\author[Murphy et al.]
{Simon J. Murphy$^{1,2}$\thanks{E-mail: simon.murphy@sydney.edu.au }\thanks{Based on data collected at the Subaru Telescope, which is operated by the National Astronomical Observatory of Japan.}, Hideyuki Saio$^{3}$, Masahide Takada-Hidai$^{4}$, Donald W. Kurtz$^{5,6}$, \and
Hiromoto Shibahashi$^{7}$, Masao Takata$^{7}$ and Daniel R. Hey$^{1,2}$
\\
$^{1}$Sydney Institute for Astronomy, School of Physics, The University of Sydney, NSW 2006, Australia\\
$^{2}$Stellar Astrophysics Centre, Department of Physics and Astronomy, Aarhus University, 8000 Aarhus C, Denmark\\
$^{3}$Astronomical Institute, Graduate School of Science, Tohoku University, Sendai  980-8578, Japan \\
$^{4}$Liberal Arts Education Center, Tokai University, Kitakaname, Hirastuka, Kanagawa 259-1292, Japan\\
$^{5}$Centre for Space Research, Physics Department, North West University, Mahikeng 2745, South Africa\\$^{6}$Jeremiah Horrocks Institute, University of Central Lancashire, Preston PR1 2HE, UK\\
$^{7}$Department of Astronomy, School of Science, The University of Tokyo, Bunkyo-ku, Tokyo 113-0033, Japan
}

\begin{document}

\maketitle

\begin{abstract}
Strong magnetic fields in chemically peculiar A-type (Ap) stars typically suppress low-overtone pressure modes (p\:modes) but allow high-overtone p\:modes to be driven. KIC\,11296437 is the first star to show both. We obtained and analysed a Subaru spectrum, from which we show that KIC\,11296437 has abundances similar to other magnetic Ap stars, and we estimate a mean magnetic field modulus of $2.8\pm0.5$\,kG. The same spectrum rules out a double-lined spectroscopic binary, and we use other techniques to rule out binarity over a wide parameter space, so the two pulsation types originate in one $\delta$\,Sct--roAp hybrid pulsator. We construct stellar models depleted in helium and demonstrate that helium settling is second to magnetic damping in suppressing low-overtone p\:modes in Ap stars. We compute the magnetic damping effect for selected p and g\:modes, and find that modes with frequencies similar to the fundamental mode are driven for polar field strengths $\lesssim4$\,kG, while other low-overtone p\:modes are driven for polar field strengths up to $\sim$1.5\,kG. We find that the high-order g\:modes commonly observed in $\gamma$\,Dor stars are heavily damped by polar fields stronger than 1--4 kG, with the damping being stronger for higher radial orders. We therefore explain the observation that no magnetic Ap stars have been observed as $\gamma$\,Dor stars. We use our helium-depleted models to calculate the $\delta$\,Sct instability strip for metallic-lined A (Am) stars, and find that driving from a Rosseland mean opacity bump at $\sim$$5\times10^4$\,K caused by the discontinuous H-ionization edge in bound-free opacity explains the observation of $\delta$\,Sct pulsations in Am stars.
\end{abstract}

\begin{keywords}
asteroseismology -- stars: magnetic fields -- stars: interiors -- stars: oscillations -- stars: variables: Delta Scuti -- stars: chemically peculiar \vspace{-4mm}
\end{keywords}

\section{Introduction}
\label{sec:intro}

Stellar pulsation frequencies are the fundamental data of asteroseismology. Magnetic fields perturb those frequencies, and can damp pulsation modes, as in the Sun, where sunspots both retard pulsation frequencies and deflect mode energy downward with some dissipation \citep{duvalletal2018}. In main-sequence stars the strongest magnetic fields are found among the chemically peculiar stars, particularly A stars, although also B and F stars; these are generically referred to as Ap stars. A rare subset of the Ap stars pulsate in high-overtone p\:modes and are known as rapidly oscillating Ap (roAp) stars \citep{kurtz1982}. Studies of a large sample of Ap stars with data from the TESS mission have shown that the roAp stars comprise only about 4 per~cent of all Ap stars, hence these stars are genuinely rare \citep{cunhaetal2019,balonaetal2019}.

The Ap stars have the strongest known global magnetic fields of any main sequence stars, with dipolar field strengths up to 50\,kG, but more typically of the order of a few kG. The lower field-strength threshold for roAp stars is not known, although we do know that the phenomenon occurs for very strong fields, with the highest known being in HD\,154708 of 24.5\,kG \citep{hubrigetal2005,hubrigetal2009,kurtzetal2006}.

\citet{saio2005} performed a non-adiabatic analysis of nonradial pulsations in the presence of a kG-strength magnetic field and found that the low-overtone p\:modes were suppressed in all models, while high-overtone p\:modes could be excited by the $\kappa$-mechanism operating in the H ionization zone, even in the presence of the strong magnetic field. This is consistent with observations of almost all known roAp stars, which generally do not show evidence of low-overtone, $\delta$\,Sct-type p-mode pulsations, and with most known Ap stars in general. KIC\,11296437 is notable as the first roAp star to pulsate simultaneously in high-overtone roAp p\:modes and low-overtone $\delta$\,Sct p\:modes, hence is of interest to test the theory and models. 

There are other stars that challenge the idea that magnetic Ap stars do not show $\delta$\,Sct pulsations. \citet{2001MNRAS.326..387K} found $\delta$\,Sct pulsations in HD\,21190, which they classified as F2III~SrEu(Si) and judged to be the most evolved Ap star known. \citet{kurtzetal2008} detected a magnetic field of $B_z = 47 \pm 13$\,G in that star, though \citet{bagnuloetal2012} reanalysed those data and concluded that the star was neither Ap nor magnetic. In a more a recent study with new spectropolarimetry, \citet{jarvinenetal2018} measured $B_z = 230 \pm 38$\,G. While these results differ by 4.6$\sigma$, the detection of a relatively weak magnetic field (for an Ap star) suggests that the suppression of low-overtone ($\delta$\,Sct) p\:modes in magnetic Ap stars may be confined to stars with stronger fields; at present we do not know the critical magnetic field strength for suppression. Other discoveries of weak magnetic fields in $\delta$\,Sct stars (\citealt{neiner&lampens2015,neineretal2017}; Zwintz et al. in press; though none are Ap stars) also point to some critical field strength for mode suppression.

\citet{kurtzetal2008} studied the visual binary HD\,218994, where both stars are in the $\delta$\,Sct instability strip. One of them is an roAp star and $\delta$\,Sct pulsations are also present. \citet{kurtzetal2008} proposed that one star was the roAp star and the other the $\delta$\,Sct star, because of the expectation that magnetic Ap stars should not show low-overtone modes, but we now recognise that it is also possible that one of the stars shows both the high-overtone roAp pulsations and the low-overtone $\delta$\,Sct pulsations. 

\citet{skarkaetal2019} discovered that HD\,99458 is a remarkable short period ($P_{\rm orb} = 2.722$\,d) eclipsing binary hosting an $\alpha^2$~CVn spotted magnetic Ap star and a low mass red dwarf companion. Binary stars among Ap stars are rare \citep{2012A&A...545A..38S}, and this is among the shortest-period examples (cf. AO\,Vel, with its 1.6-d period \citealt{gonzalezetal2006}). Interestingly in the present context, \citet{skarkaetal2019} also found that HD\,99458 is a multi-periodic $\delta$\,Sct star. Similarly, \citet{escorzaetal2016} discovered $\delta$\,Sct pulsations in the Ap star HD\,41641. Thus here are two more cases of low-overtone $\delta$\,Sct pulsation in Ap stars, but no high-overtone roAp pulsations are reported for either. The magnetic field strengths of the Ap primaries in these two cases are not yet known.

The high-overtone pulsations of roAp stars have periods in the range of 4--24\,min. The strength and geometry of their global, approximately dipolar magnetic fields constrain the pulsation modes. The roAp stars are oblique pulsators: their nonradial pulsations have axes that are inclined to the rotation axis of the star and nearly aligned with the magnetic axis \citep{kurtz1982,shibahashitakata1993,takatashiba94,takatashiba95,saio&gautschy2004,bigot&kurtz2011}. 

Magnetic Ap stars on average rotate much more slowly than normal A stars, with rotation periods that can be years or even decades \citep{mathysetal2020}. The mechanism of the rotational braking, probably magnetic  \citep{stepien2000}, is still uncertain. Without substantial surface convection zones, atomic diffusion becomes an important process governing the observed abundances of A stars \citep{michaud1970,vauclairetal1974,theadoetal2012,dealetal2016}. Those that rotate rapidly, usually with velocities exceeding 150\,km\,s$^{-1}$ \citep{zorec&royer2012}, are mixed by rotation; the Ap stars are not, allowing certain elements to accumulate in the stellar atmosphere while others (such as helium) gravitationally settle. In this sense, Ap stars are similar to the non-magnetic Am stars, with the key difference that the magnetic fields of Ap stars concentrate peculiarities near the magnetic poles.

The driving of low-overtone pulsations persists in Am stars, despite the gravitational settling of helium, albeit in a narrower instability strip and at lower incidence than for normal stars \citep{vauclair1976,kurtz1989,smalleyetal2011,smalleyetal2017}. It appears that turbulent pressure contributes to pulsational driving in these stars \citep{houdek2000,antocietal2014,antocietal2019}, offering one explanation why some of them pulsate. The suppression of convection in roAp stars by their magnetic fields further explains why these stars have not been observed with low-overtone p\:modes. Until now.

Four years of \textit{Kepler} observations for KIC\,11296437 are available in long-cadence (LC) mode, with 29.5-min sampling. \citet{sna2013} have shown that barycentric corrections made to the timing of {\it Kepler} observations modulate the exactly regular sampling of the data. Pulsation cycles are thus sampled at different phases across the satellite's orbit of the Sun. A consequence is that Nyquist aliases are split into multiplets that can be identified by their shape. Real pulsation frequencies are distinguishable from these aliases and their frequencies are completely recoverable, even in the super-Nyquist regime, that is, when the sampling interval is longer than half the pulsation period.

This has opened up the study of hundreds of stars that are high frequency pulsators, but for which only LC {\it Kepler} data are available. \citet{heyetal2019} exploited this to discover six new rapidly oscillating Ap (roAp) stars in the {\it Kepler} LC data. \citet{cunhaetal2019} discovered a further 5 new roAp stars in the first two sectors of TESS observations, and \citet{balonaetal2019} found more in TESS sectors 1--7. The number of known roAp stars is growing, particularly because of the TESS mission, and the census now gives of the order of 100. 

\begin{figure}
\centering	
\includegraphics[width=1.0\linewidth]{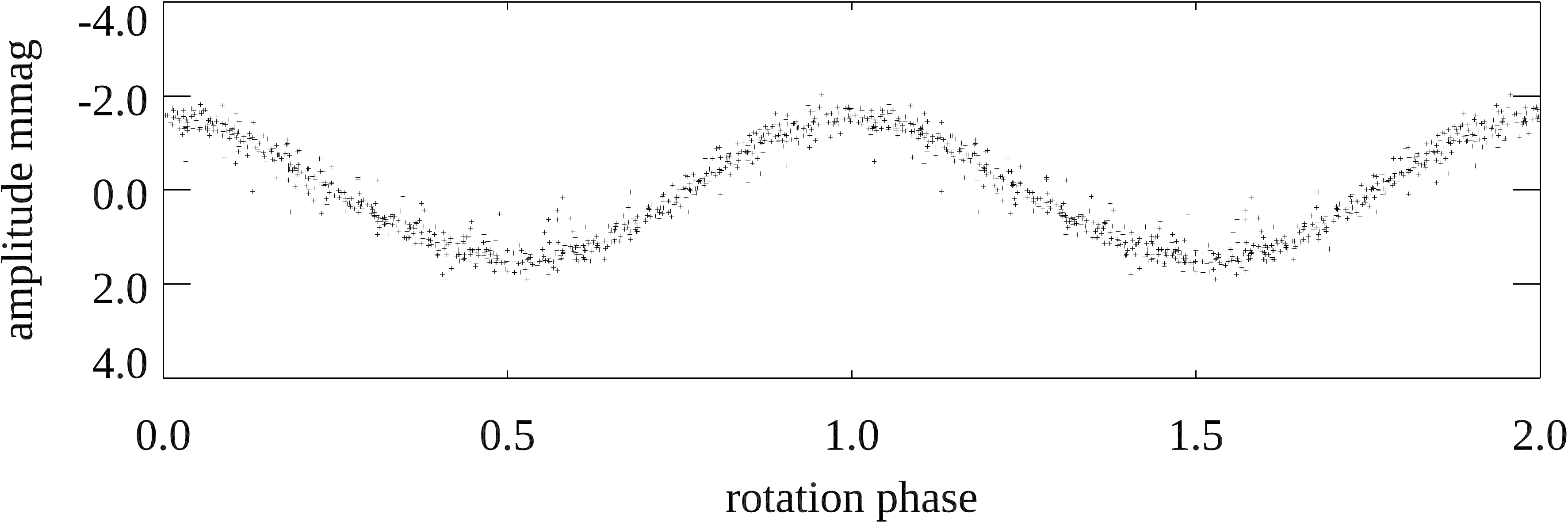}
\includegraphics[width=1.0\linewidth]{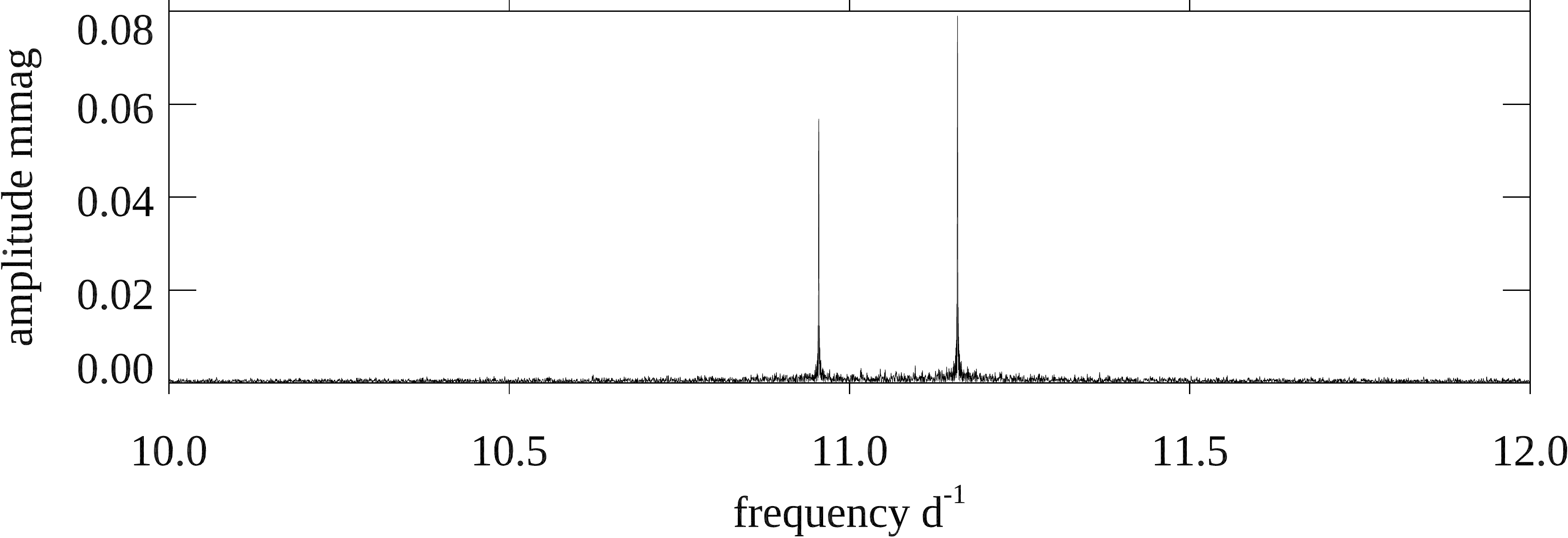}
\caption{{\bf Top}: The rotational variation of KIC\,11296437, phased on the period 7.12433\,d, shown for two rotation cycles. The long cadence data have been binned by a factor of 10, so each data point represents 4.9\,hr. It is clear that there is only one hump per rotation, which is typical of Ap stars where only one magnetic pole, hence usually one set of spots, is visible over the rotation. {\bf Bottom}: an amplitude spectrum of the data in a low frequency range after the rotational variations have been removed. The two low-overtone p-mode frequencies are clear.}
\label{fig:intro}
\end{figure}

KIC\,11296437 has $T_{\rm  eff} \approx 7000$\,K, typical of the roAp stars. Using {\it Kepler} long cadence data, \citet{heyetal2019} discovered super-Nyquist roAp pulsations in this star, and also showed low-overtone p\:modes. In the roAp star frequency range, KIC\,11296437 shows a triplet from an obliquely pulsating dipole mode with frequencies near 121\,d$^{-1}$ (1.4\,mHz), i.e., periods near 12\,min. These are unambiguously the correct frequencies using the super-Nyquist technique. \citet{heyetal2019} used the frequency multiplet to constrain the pulsation geometry for this star and showed that $i + \beta \le 90^\circ$, where $i$ is the rotational inclination and $\beta$ is the magnetic obliquity. That constraint on the pulsation geometry shows that only one pulsation pole is visible over the rotation cycle of this star, and that is consistent with the single-hump rotational spot variation of the star over the 7.12433-d rotation period seen in the top panel of Fig.\,\ref{fig:intro}. That also confirms that this star is an $\alpha^2$~CVn spotted magnetic Ap star. 

KIC\,11296437 also clearly shows low-overtone p\:modes with frequencies of 10.95 and 11.15\,d$^{-1}$, which are illustrated in the bottom panel of Fig.\,\ref{fig:intro}. We make a strong case in Sec.\,\ref{sec:binary} that the star is not binary, over a wide range of parameter space, so the high- and low-overtone pulsations originate in the same star. Thus KIC\,11296437 is important to our understanding of pulsation driving and damping in roAp stars. This star presents two interesting questions: Are there weakly magnetic roAp stars that can also pulsate in low-overtone $\delta$\,Sct p\:modes, and is KIC\,11296437 an example? If so, we should find others. Or is KIC\,11296437 a strongly magnetic roAp star that demonstrates that low-overtone p\:modes are not necessarily damped by the strong magnetic field? In that case, the theory of \citet{saio2005} requires modification. Both of these may improve our understanding of the interaction of pulsation in the presence of significant global magnetic fields. 

A third question is to what extent the damping of low-order pulsations is dependent on the magnetic field strength. Such pulsations are primarily driven by the $\kappa$-mechanism operating on the second partial ionization zone of helium, but if gravitational settling of helium drains this source of opacity, are low-order pulsations damped in these slow rotators regardless of the field strength?

To address these questions we obtained a high resolution spectrum of KIC\,11296437 to look for evidence of a magnetic field and to perform an abundance analysis to look for Ap-type abundance anomalies. This paper presents the results of that analysis. We also calculate stellar pulsation models of varying magnetic field strengths and helium depletions, and look for other possible explanations for the observed low-frequency pulsations.


\section{Spectroscopic observations and data reduction}  
\label{sec:observations}

KIC\,11296437 was observed with the High Dispersion Spectrograph  \citep[HDS;][]{2002PASJ...54..855N} on the Subaru telescope on 2016 November 16. An \'echelle spectrum was obtained in a standard StdYc setup covering the wavelength range of 4380--7110\,\AA\ with an exposure time of 2400\,s, and using the image slicer \#3 with $1\times1$ binning, which yields a wavelength resolution of 160\,000. Basic data and the atmospheric parameters for the star are shown in Table\:\ref{tab:01}.

Standard data reduction procedures (bias subtraction, background subtraction, cosmic ray removal, flat-fielding, extraction of 1D spectra, wavelength calibration, and normalisation) were carried out with the {\sc iraf} \'echelle package\footnote{{\sc iraf} is distributed by the National Optical Astronomy Observatory, which is operated by the Association of Universities for Research in Astronomy, Inc., under cooperative agreement with the National Science Foundation.}. The resultant spectrum consists of a blue part with wavelength coverage of 4380--5700\,\AA, and a red part with 5785--7110\,\AA.  Signal-to-noise ratios (S/N) were measured on peak continua of each order, and found to be 50--80 for the blue part and 75--80 for the red part. To make a Doppler correction, the radial velocity was measured using 50 Fe~{\sc i} and 17 Fe~{\sc ii} lines adopted from \citet{2000ApJ...530..783W}. The radial velocity, averaged by weights of numbers of lines, is $V_{\rm r} = -22.3 \pm 0.5$\,km\,s$^{-1}$, giving a heliocentric radial velocity of  $V_{\rm h} = - 33.0 \pm 0.5$\,km\,s$^{-1}$, as listed in Table\:\ref{tab:01}.

\begin{table}
\caption{Fundamental data and model parameters of KIC\,11296437. }
\centering
\begin{tabular}{lr}
\toprule
 \multicolumn{2}{c}{Position and magnitude} \\
 \midrule
 Right Ascension (J2000.0)    & $19^{\rm h} 26^{\rm m} 18^{\rm s}$ \\
 Declination  (J2000.0)       & $+49\degr 05\arcmin 52\arcsec$\\
  Kp (mag)${}^{*}$       & 11.74 \\
\midrule
 \multicolumn{2}{c}{Literature parameters} \\
\midrule
$T_{\rm eff}$ (K)    & $7000 \pm 300$ \\
$\log g$ (cgs)        & $3.9 \pm 0.1$ \\
 $[{\rm Fe/H}] $     & $-0.1 \pm 0.2$  \\
 $\log_{10} (L/{\rm L_{\odot}})^{\dagger}$ & $1.105\pm0.027$ \\
\midrule
 \multicolumn{2}{c}{Spectroscopically derived parameters (Sec.\,\ref{ssec:hires})} \\
\midrule
$T_{\rm eff}$ (K)    & $7450 \pm 300$ \\
 $\log g$ (cgs)        & $4.34 \pm 0.10$  \\
 $\xi$ (km\,s$^{-1}$)& $0.8\phantom{0} \pm 0.3\phantom{0}$ \\
 $[{\rm Fe/H}] $      & $-0.03 \pm 0.16$  \\
$v\sin i$ (km\,s$^{-1}$)         & \phantom{000}$<$2.38\phantom{000}   \\
 $V_{\rm h}$ (km\,s$^{-1}$)         & $-33.0 \pm 0.5\phantom{0}$   \\
\midrule
 \multicolumn{2}{c}{Asteroseismically derived parameters} \\
  \multicolumn{2}{c}{(Best-fitting model; Sec.\,\ref{sec:seismo})} \\ 
\midrule
$M$/M$_\odot$   & 1.75 \\
$X$ (H fraction)    & 0.70\\
$Y$ (He fraction)    & 0.28 \\
$Z$ (metal fraction)   & 0.02 \\
$T_{\rm eff}$ & 6930 \\
$\log L/{\rm L}_{\odot}$ & 1.06 \\
\bottomrule
\end{tabular}
{\footnotesize \\
\vspace{1mm}$^{*}$Kp is the white light $\it Kepler$ magnitude taken from the revised {\it Kepler} Input Catalog (KIC) \citep{Huber2014}. $^{\dagger}$Luminosity calculated using these literature parameters and Gaia DR2 data (Sec.\,\ref{sec:atmosphere}).}
\label{tab:01}
\end{table}

\section{Atmospheric parameters}
\label{sec:atmosphere}

The atmospheric parameters of magnetic Ap stars are difficult to determine because the atmosphere is heavily stratified. The accumulation of rare earth elements high in the atmosphere blocks flux, particularly in the UV where these elements have many absorption lines, leading to steep temperature gradients in each of the line-forming layers of these elements. This is known as the line-blocking effect, and it is analogous to the greenhouse effect in Earth's atmosphere, with the result being that the blocked flux heats the lower layers of the atmosphere. In the case of cool Ap stars such as KIC\,11296437, the spectral energy distributions shift to redder wavelengths.

The temperature gradients of the atmosphere are changed, becoming steeper for the higher atmosphere where the lines of rare earth elements are formed in layers cooler than normal, and gradients are shallower in the warmed lower atmosphere. The deduced $T_{\rm eff}$ of the star therefore has considerable dependence on the lines used for the analysis. Naturally, we arrived at different effective temperatures according to the methods employed: from the ionization equilibrium of Fe\,{\sc i} lines in the Subaru spectrum we arrived at $T_{\rm eff}= 7450$\,K; an analysis of the star's Balmer lines gave $T_{\rm eff}=7230$\,K (H$\beta$) and 7030\,K (H$\alpha$), and fitting of spectral energy distributions (SEDs) yielded $T_{\rm eff} = 7000$\,K. We describe these analyses in this section (\ref{ssec:balmer}--\ref{ssec:hires}). Ultimately, our abundance analysis was carried out at 7450\,K and separately at 7000\,K (Sec.\,\ref{sec:abundances}).

Stellar parameters are also available in the literature. We used these as starting points for our analysis. The effective temperature, $T_{\rm eff} = 7000 \pm 300$\,K, was obtained from LAMOST DR4 spectroscopy \citep{zhaoetal2012b}\footnote{\url{http://dr4.lamost.org/search/} with obsid = 249608193 for KIC\,11296437} and is based on SED fitting \citep{wuetal2011}. The LAMOST survey also gives a metallicity $[{\rm Fe/H}]$\footnote{$[{\rm X/Y}] \equiv \log ({\rm X/Y})_{\rm star} - \log ({\rm X/Y})_{\sun}$ is used throughout the paper.} that we rounded to $[{\rm Fe/H}] = -0.10 \pm 0.20$, with an assumed uncertainty.

We calculated the distance to KIC\,11296437 using the Gaia DR2 parallax \citep{gaiacollaboration2018a} and the normalized posterior distribution and length-scale model of \citet{bailer-jonesetal2018}. This produced a distribution of distances that a Monte Carlo process could sample, the median of which was 807\,pc. We calculated the luminosity using a similar method to \citet{murphyetal2019}, except that we propagated the uncertainties of temperature, surface gravity and metallicity into the bolometric corrections provided by the MIST tables \citep{choietal2016,dotter2016}. This makes the luminosity calculated here more robust than that of \citet{murphyetal2019} and \citet{heyetal2019}. We computed the distribution of bolometric corrections via a Monte Carlo process, using the SDSS $g$ magnitude of \mbox{$11.822\pm0.020$\,mag}, $T_{\rm eff} = 7000\pm300$\,K, [Fe/H]$ = -0.1\pm0.2$ and $\log g = 3.9\pm0.1$ as inputs. Small changes in these inputs do not strongly affect the derived luminosity -- a 4$\sigma$ change in $\log g$ changes $\log L$ by only 0.01\,dex, which is much less than 1$\sigma$ in the luminosity. We used a three-dimensional map of Galactic reddening \citep{greenetal2018} and the python {\tt dustmaps} package \citep{green2018} to sample extinctions, and computed a log luminosity of $\log_{10} (L/{\rm L_{\odot}}) = 1.105\pm0.027$ for KIC\,11296437. The above luminosity and temperature suggest a mass of $M = 1.75\pm0.25$\,M$_{\odot}$ and a corresponding $\log g$ of $3.88\pm0.10$\,dex, compatible with, but not dominated by, the inputs used for the bolometric correction. If the parameters from high-resolution spectroscopy (Sec.\,\ref{ssec:hires}) are used as inputs instead, the luminosity changes by less than 1$\sigma$.

\subsection{Balmer line analysis}
\label{ssec:balmer}

\begin{figure}
\centering
\includegraphics[width=0.95\columnwidth]{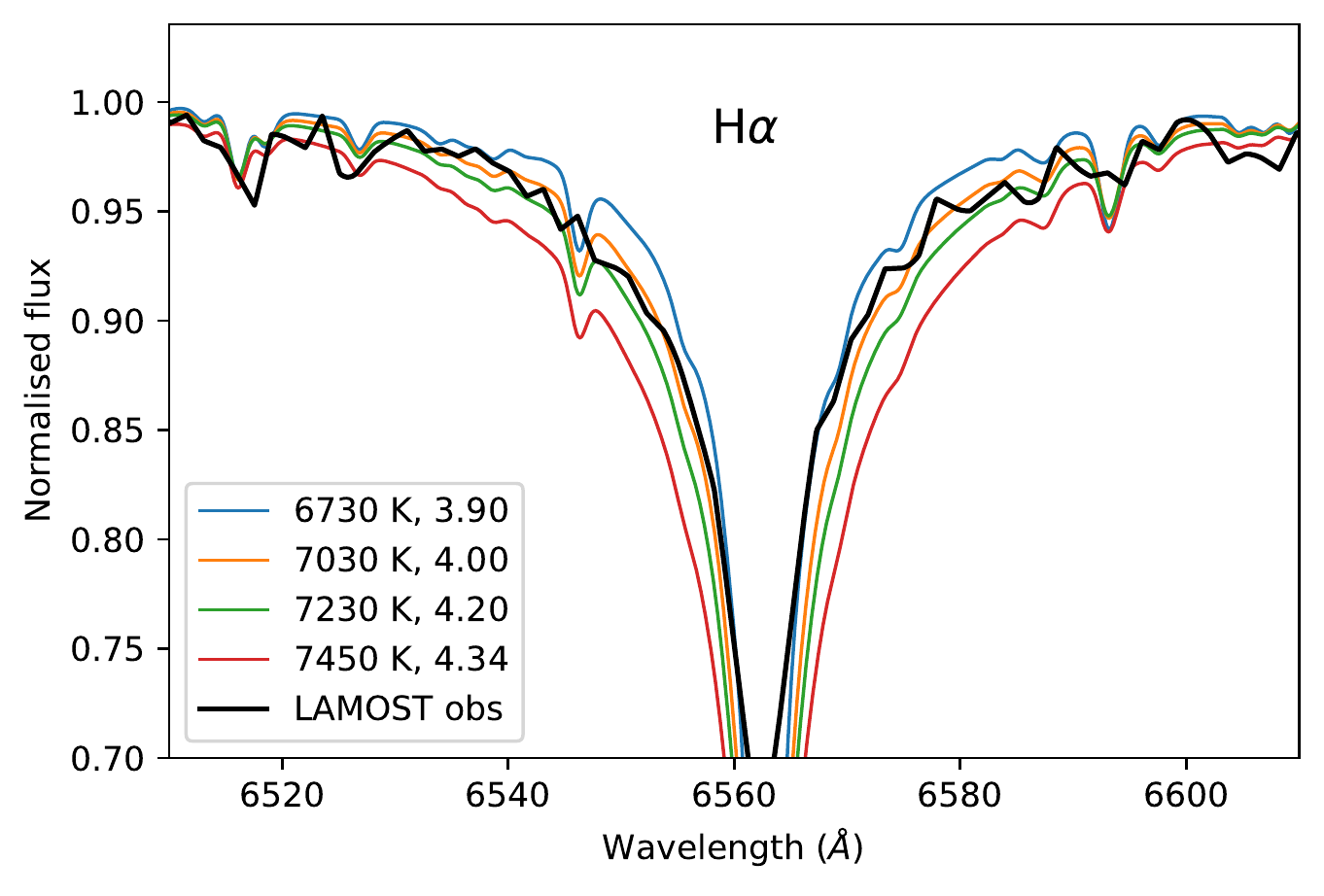}
\includegraphics[width=0.95\columnwidth]{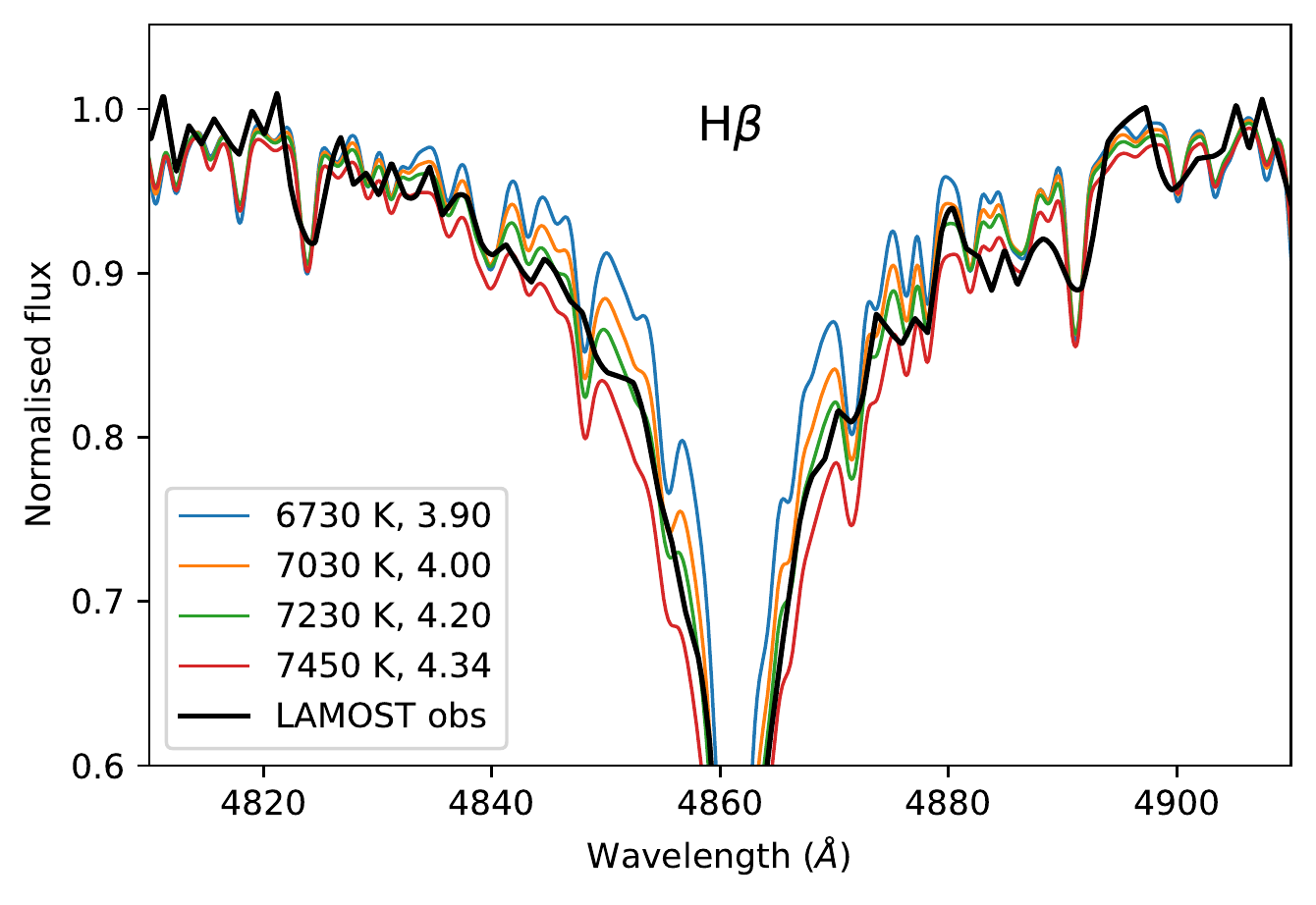}
\caption{Observed profiles of H$\alpha$ (top) and H$\beta$ (bottom) from the LAMOST spectrum, compared to a subset of analysed model spectra spanning a small range of parameter space in $T_{\rm eff}$ and $\log g$ but fixed [Fe/H] of $-0.1$\,dex. Also shown (red) is the best-fitting model derived using the ionization equilibrium of Fe\,{\sc i} lines (Sec.\,\ref{ssec:hires}) in the Subaru spectrum, with $T_{\rm eff} = 7450$\,K and $\log g =4.34$.}
\label{fig:balmer}
\end{figure}

For late A and early F stars, $\log g$ is particularly difficult to infer from spectra because of degeneracies with metallicity and temperature. Fortunately, the Gaia parallax constrains the luminosity of KIC\,11296437 well, which in turn constrains $\log g$. Using the assumed metallicity of [Fe/H] = $-0.1$ and restricting $\log g$ to lie close to the calculated value, we compared model profiles of the H$\beta$ and H$\alpha$ regions to the low-resolution LAMOST spectrum to infer $T_{\rm eff}$ (Fig.\,\ref{fig:balmer}). The Subaru spectrum could not be used for this because edges of the \'echelle orders fall close to the Balmer-line wings, causing heavy dependence on normalization. We found that the best fitting temperature differs between H$\alpha$ and H$\beta$, but both are consistent with $\sim$7000\,K, within the systematic uncertainties of normalisation. We therefore confirm the $T_{\rm eff}$ value inferred from the LAMOST spectrum given in the LAMOST DR4 catalogue.

\subsection{Parameters from the high-resolution spectrum}
\label{ssec:hires}

We determined the stellar atmospheric parameters using the iterative procedures outlined by \citet{takedaetal2002}. The equivalent widths $W_{\lambda}$ and abundances determined from each of the selected Fe\,{\sc i}~and~{\sc ii} lines were used to determine $T_{\rm eff}$, $\log g$, and microturbulence $\xi$, by perturbing those three parameters from their initial values.

The first step in each iteration is to calculate the microturbulence, $\xi$, based on an {\small ATLAS9} 1D LTE model atmosphere \citep{1993KurCD..13.....K} constructed with the current $T_{\rm eff}$, $\log g$, and [Fe/H], by requiring that the abundances derived from 50  Fe~{\sc i} lines are independent of their corresponding equivalent widths, $W_{\lambda}$. In this procedure we used the {\sc sptool} software package developed by Y. Takeda.\footnote{\url{http://optik2.mtk.nao.ac.jp/~takeda/sptool/}, which is based on Kurucz's {\small ATLAS9} 1D LTE model atmospheres and {\small WIDTH9} program for abundance analysis \citep{1993KurCD..13.....K}. }
With the new $\xi$, [Fe/H] was redetermined as an average of the Fe\,{\sc i} and {\sc ii} abundances, weighted by the number of lines. 
The excitation equilibrium of Fe\,{\sc i} lines gives $T_{\rm eff}$ by requiring that the Fe\,{\sc i} abundance is independent of the lower excitation potential, $\chi$.
Finally, $\log g$ was evaluated under the requirement of ionization equilibrium, such that the Fe abundances derived from Fe\,{\sc i} and {\sc ii} lines are equal. The parameters thereby determined are shown in Table\:\ref{tab:01}. We note that the $\log g$ value determined in this way differs from the value derived via the stellar parallax; the two are independent measurements from different data.

We measured the rotational line broadening, $v\sin i$, using the same procedure applied to the extremely slowly rotating A star KIC\,11145123 \citep{takada-hidaietal2017}, namely that of \citet{takedaetal2008}, in which $v\sin i$ is obtained as the intercept of a least-squares fit of the line broadening, $v_{\rm rm}$, as a function of $W_{\lambda}$. For this, we used 55 \ion{Fe}{i} lines that we considered to be unblended, to be consistent with the Fe abundance of $7.5\pm0.3$\,dex, and to be symmetric. We assumed 10\:per\:cent uncertainties on $W_{\lambda}$. We measured $v_{\rm rm} = 2.38\pm0.12$\,km\,s$^{-1}$, which includes both rotational and macroturbulent broadening.\footnote{Using eq.\,5 of \citet{kjeldsen&bedding1995}, we calculate that the two low-overtone modes contribute a negligible $\sim$5\,m\,s$^{-1}$ of pulsational broadening.} As such, we determine an upper limit on $v\sin i$ of 2.38\,km\,s$^{-1}$.

We use this limit on $v\sin i$ to estimate a limit on the inclination angle between the stellar rotation axis and the line of sight, $i$, for later use. With the stellar luminosity and the temperature with which it was calculated, we determine a stellar radius of
\begin{eqnarray}
R / {\rm R}_{\odot} = \sqrt{\frac{L/{\rm L}_{\odot}}{(T_{\rm eff}/{\rm T}_{\rm eff,Sun})^4}} = \sqrt{\frac{10^{1.105}}{(7000/5777)^4}} = 2.43.
\end{eqnarray}
The circumference of the star is then $2\uppi R = 1.06\times10^7$\,km. For a rotation period of $7.12$\,d, we calculate an equatorial rotation velocity, $v_{\rm eq} = 17.3$\,km\,s$^{-1}$, and the limit on the stellar inclination given by $\sin i = v\sin i / v_{\rm eq} \leq 0.138$, is $i<7.9^{\circ}\pm0.8^{\circ}$. That is, the star is seen almost pole-on.

\subsection{Magnetic field}
\label{ssec:mag}

We searched for a magnetic field by applying the method of \citet{mathys1990}, which uses the ratio of the equivalent widths ($W_{\lambda}$) of the Fe~{\sc ii} lines at 6147.7\,\AA\  ($W_{\lambda} = 48.1$\,m\AA) and 6149.2\,\AA\ ($W_{\lambda} = 39.0$\,m\AA) to estimate the mean magnetic field modulus $\langle H \rangle$.
Since KIC\,11296437 is a cool Ap star, the equivalent width of the Fe~{\sc ii} 6147.7 line is substantially enhanced by the neighbouring Fe~{\sc i} 6147.8 line \citep{mathys&lanz1992}. We therefore modelled the individual contributions from the Fe~{\sc i} and Fe~{\sc ii} components simultaneously, using the best-fitting spectroscopic parameters (Table\:\ref{tab:01}), and found $W_{\rm Fe~\small{II}~6147.7} = 43.3$\,m\AA. Using only $W_{\lambda}$ for the two Fe~{\sc ii} lines, whose equivalent widths differ by $\delta W_{\lambda} = 43.3-39.0=4.3$\,m\AA\ and whose mean width is $\overline{W_{\lambda}}=(43.3-39.0)/2=41.2$\,m\AA, we calculated the ratio $R=\delta W_{\lambda}/\overline{W_{\lambda}}=4.3/41.2=0.104\pm0.010$, which we apply to the empirical relation of \citet{mathys&lanz1992} to arrive at $\langle H \rangle = 2.8\pm0.5$\,kG. The uncertainty encompasses our measurement uncertainty (10\:per\:cent) and the scatter in the empirical relation. Since no Zeeman splitting is observed, our abundance analysis (Sec.\,\ref{sec:abundances}) does not account for any magnetic broadening.

The quoted validity of the relation in \citet{mathys&lanz1992} is for $\langle H \rangle$ of 3--5\,kG and spectral types not later than A6, and those authors recommend that their relation not be used blindly. The spectral type constraint is associated with the blending of the Fe~{\sc i} and Fe~{\sc ii} lines, which we have handled via simultaneous measurement, and we note that our estimate of $\langle H \rangle$ is only just outside the quoted range. We also note that similar ratios $R$ have been measured for the late-type Ap stars HD\,24712 (A9p, SrCrEu) and HD\,176232 (10 Aql,  A6p Sr), with $R = 0.103$ and 0.124, respectively \citep{mathys&lanz1992}. These stars both have $\langle H \rangle \sim 2$\,kG \citep{preston1971,preston1972,ryabchikovaetal2004}, depending on the rotation phase, and like KIC\,11296437, neither shows Zeeman splitting. Hence, in the absence of spectropolarimetry, our estimate of $\langle H \rangle$ appears justified.

\begin{figure}
\centering
\includegraphics[width=0.95\columnwidth]{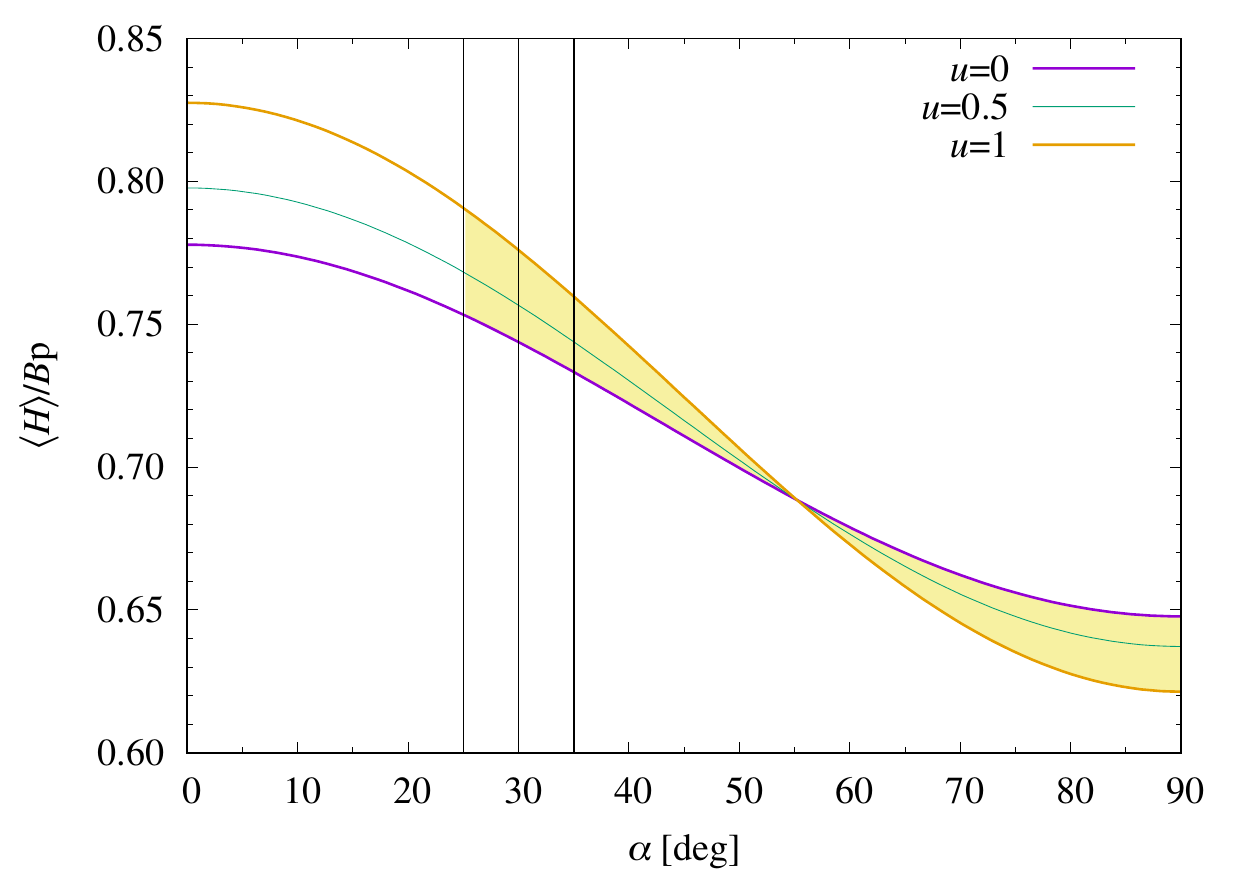}
\caption{The ratio of the mean magnetic field modulus $\langle H \rangle$ and the polar field strength, $B$p, as a function of the inclination of the magnetic axis, $\alpha$, is shown as the shaded yellow region. Curves for three values of the limb darkening parameter, $u$, are shown, and the calculated limit on $\alpha$ ($>30\pm5^{\circ}$) is shown as a vertical black line.}
\label{fig:b}
\end{figure}

Assuming a simple dipolar configuration and knowing the rotational inclination, $i$, it is possible to estimate the value of the polar field strength, $B{\rm p}$, from $\langle H \rangle$. As described by \citet{preston1969}, the value depends on the adopted limb-darkening coefficient, $u$, the angle between the magnetic axis and the line of sight, $\alpha$, and the angle between the rotational and magnetic axes which is designated $\beta$. From the constraint from \citet{heyetal2019} that $\tan i \tan \beta = 0.11\pm0.01$ and $i<7.9\pm0.8^{\circ}$ (Sec.\,\ref{ssec:hires}), we infer that $\beta>38\pm4^{\circ}$. Since $\alpha$ lies in the range $\beta - i < \alpha < \beta + i$, we determine a lower limit of $\alpha > 30\pm5^{\circ}$. Fig.\,\ref{fig:b} shows the corresponding allowed ratio of field strengths, $\langle H \rangle / B$p. Adopting $u=0.5$, we determine that $0.64 < \langle H \rangle/B{\rm p} < 0.76\pm0.01$, hence $3.7\pm0.7$ $< B{\rm p/kG} <$ $4.4\pm0.8$.

\section{Abundance Analysis}
\label{sec:abundances}

\subsection{Abundance results}
\label{sec.results}

We determined the stellar abundances twice using the Subaru spectrum and two different model atmospheres. One analysis used $T_{\rm eff} = 7000$\,K, $\log g$ = 3.9, while the other used the parameters determined from the high-resolution spectrum itself, $T_{\rm eff} = 7450$\,K and $\log g = 4.34$. The difference in inferred abundances indicates the systematic uncertainty on the analysis.

We used the {\sc spshow} program in the {\sc sptool} software package to measure $W_{\lambda}$ for lines of selected elements by Gaussian profile fitting in most cases, but direct integration in others. The program {\sc width} was used to derive abundances from  $W_{\lambda}$. Atomic data of wavelengths, excitation potentials and $\log gf$ values are based on \citet{1995KurCD..23.....K} and \citet{2004A&A...425..263C} (for updated 
$\log gf$ values), but when $\log gf$ values are available from the NIST database\footnote{\url{https://www.nist.go/pml/atomic-spectra-database/}}, those were adopted instead. For the $\log gf$ values of 34 Fe~{\sc ii} lines, 17 were adopted from \citet{2000ApJ...530..783W} and the other 17 from \citet{2004A&A...425..263C}. The derived abundances for each element and ion are summarised for both model atmospheres in Table\,\ref{tab:abundances}.

\begin{table*}
\caption{Summary of abundances of each element and ion. For each ion of element $X$ (column 1), average abundances ($\log X$) are given in column 2, and expressed with respect to H and Fe in columns 3 and 4.
Standard deviations of $\log X$ are shown in column 5, based on the number of lines measured (column 6).
The numbers of doublet or triplet lines, and lines of hyperfine structure (hfs) fitted to derive the abundances are given in column 7. Uncertain abundances are noted as ``uncertain'' in the remarks. Solar abundances from \citet{asplundetal2009} are listed in column 8 for reference. Column 9 indicates the abundances for the $T_{\rm eff}$=7000\,K model, and in column 10 we give the difference from the 7450\,K model ($\Delta \log X = \log X_{7450} - \log X_{7000}$). We use this as the systematic uncertainty, which is combined in quadrature with the standard deviation to give the total uncertainty in the final column.}
\begin{center}
\begin{tabular}{HlrrrcHHrcrrrr}
\toprule
\multicolumn{2}{c}{} & \multicolumn{4}{c}{7450\,K model} & \multicolumn{2}{c}{} & \multicolumn{3}{c}{} & \multicolumn{2}{c}{7000\,K model} & \\
\cmidrule(lr){12-13}
\cmidrule(lr){3-7}
Code & Ion & $\log X$ & [$X$/H] & [$X$/Fe] & Std.Dev. & Error & Total & No. of & Remark & Sun & [$X$/H] & $\Delta \log X$ & Total\\
 & & average & & & 1$\sigma$ & Tg$\xi$ & error & lines & & & & & unc.\\
\midrule
 3.00 & Li~{\sc i} & $\le$ 1.46 & $\le$ 0.41 & $\le$ 0.44 & $\cdots$ & $\cdots$ & $\cdots$ & 2 & 1 single line & 1.05 & $\le$ 0.15 & $\cdots$ & $\cdots$ \\
 6.00 & C~{\sc i} & 8.54 & 0.11 & 0.14 & 0.16 & 0.08 & 0.18 & 22 & 1 doublet, 1 triplet & 8.43 & 0.13 & $-$0.02 & 0.16 \\
 8.00 & O~{\sc i} & 8.24 & $-$0.45 & $-$0.42 & 0.08 & 0.24 & 0.25 & 3 & 3 triplets & 8.69 & $$-$$0.36 & $-$0.09 & 0.12 \\
11.00 & Na~{\sc i} & 6.52 & 0.28 & 0.31 & 0.09 & 0.07 & 0.11 & 6 & 2 single lines & 6.24 & 0.13 & 0.15 & 0.17 \\
12.00 & Mg~{\sc i} & 8.03 & 0.43 & 0.46 & 0.07 & 0.08 & 0.11 & 5 & & 7.60 & 0.26 & 0.17 & 0.18 \\
12.01 & Mg~{\sc ii} & 7.71 & 0.11 & 0.14 & $\cdots$ & 0.20 & 0.20 & 1 & 1 triplet & 7.60 & 0.19 & $-$0.08 & 0.08 \\
13.00 & Al~{\sc i} & 7.00 & 0.55 & 0.58 & 0.15 & 0.06 & 0.16 & 3 & & 6.45 & 0.42 & 0.13 & 0.20 \\
13.01 & Al~{\sc ii} & 7.23 & 0.78 & 0.81 & 0.20 & 0.15 & 0.25 & 3 & uncertain & 6.45 & 0.94 & $-$0.16 & 0.26 \\
14.00 & Si~{\sc i} & 8.05 & 0.54 & 0.57 & 0.23 & 0.05 & 0.24 & 37 & & 7.51 & 0.44 & 0.10 & 0.25 \\
14.01 & Si~{\sc ii} & 8.05 & 0.54 & 0.57 & 0.12 & 0.16 & 0.20 & 4 & & 7.51 & 0.70 & $-$0.16 & 0.20 \\
16.00 & S~{\sc i} & 7.11 & $-$0.01 & 0.02 & 0.13 & 0.07 & 0.15 & 8 & 1 doublet, 5 triplets & 7.12 & $$-$$0.07 & 0.06 & 0.14 \\
19.00 & K~{\sc i} & 5.66 & 0.63 & 0.66 & $\cdots$ & 0.06 & 0.06 & 1 & uncertain & 5.03 & 0.50 & 0.13 & 0.13 \\
20.00 & Ca~{\sc i} & 6.46 & 0.12 & 0.15 & 0.17 & 0.09 & 0.19 & 20 & & 6.34 & $$-$$0.09 & 0.21 & 0.27 \\
20.01 & Ca~{\sc ii} & 6.58 & 0.24 & 0.27 & 0.06 & 0.08 & 0.10 & 4 & 1 single line & 6.34 & 0.25 & $-$0.01 & 0.06 \\
21.01 & Sc~{\sc ii} & 2.92 & $-$0.23 & $-$0.20 & 0.24 & 0.17 & 0.29 & 9 & 8 hfsfit & 3.15 & $$-$$0.51 & 0.28 & 0.37 \\
22.00 & Ti~{\sc i} & 5.43 & 0.48 & 0.51 & 0.16 & 0.12 & 0.20 & 38 & 27 hfsfit & 4.95 & 0.19 & 0.29 & 0.33 \\
22.01 & Ti~{\sc ii} & 5.55 & 0.60 & 0.63 & 0.15 & 0.09 & 0.17 & 33 & & 4.95 & 0.28 & 0.32 & 0.35 \\
23.00 & V~{\sc i} & 4.78 & 0.85 & 0.88 & 0.11 & 0.14 & 0.18 & 13 & 7 hfsfit & 3.93 & 0.52 & 0.33 & 0.35 \\
23.01 & V~{\sc ii} & 4.65 & 0.72 & 0.75 & 0.13 & 0.05 & 0.14 & 11 & & 3.93 & 0.49 & 0.23 & 0.26 \\
24.00 & Cr~{\sc i} & 6.34 & 0.70 & 0.73 & 0.10 & 0.10 & 0.14 & 46 & & 5.64 & 0.45 & 0.25 & 0.27 \\
24.01 & Cr~{\sc ii} & 6.17 & 0.53 & 0.56 & 0.17 & 0.09 & 0.19 & 23 & & 5.64 & 0.30 & 0.23 & 0.29 \\
25.00 & Mn~{\sc i} & 6.34 & 0.91 & 0.94 & 0.18 & 0.11 & 0.21 & 27 & 27 hfsfit & 5.43 & 0.66 & 0.25 & 0.31 \\
25.01 & Mn~{\sc ii} & 6.38 & 0.95 & 0.98 & 0.14 & 0.07 & 0.16 & 9 & 9 hfsfit & 5.43 & 0.87 & 0.08 & 0.16 \\
26.00 & Fe~{\sc i} & 7.47 & $-$0.03 & 0.00 & 0.10 & 0.14 & 0.17 & 50 & & 7.50 & $$-$$0.35 & 0.32 & 0.34 \\
26.01 & Fe~{\sc ii} & 7.47 & $-$0.03 & 0.00 & 0.13 & 0.07 & 0.15 & 34 & & 7.50 & $$-$$0.23 & 0.20 & 0.24 \\
27.00 & Co~{\sc i} & 6.61 & 1.62 & 1.65 & 0.20 & 0.10 & 0.22 & 72 & 72 hfsfit & 4.99 & 1.39 & 0.23 & 0.30 \\
27.01 & Co~{\sc ii} & 6.87 & 1.88 & 1.91 & 0.09 & 0.07 & 0.11 & 3 & & 4.99 & 1.68 & 0.20 & 0.22 \\
28.00 & Ni~{\sc i} & 6.22 & 0.00 & 0.03 & 0.17 & 0.09 & 0.19 & 57 & & 6.22 & $$-$$0.22 & 0.22 & 0.28 \\
29.00 & Cu~{\sc i} & 3.84 & $-$0.35 & $-$0.32 & 0.07 & 0.11 & 0.13 & 2 & 2 hfsfit & 4.19 & $$-$$0.69 & 0.34 & 0.35 \\
30.00 & Zn~{\sc i} & 4.20 & $-$0.36 & $-$0.33 & 0.08 & 0.10 & 0.13 & 3 & & 4.56 & $$-$$0.61 & 0.25 & 0.26 \\
38.00 & Sr~{\sc i} & 3.59 & 0.72 & 0.75 & $\cdots$ & 0.14 & 0.14 & 1 & & 2.87 & 0.40 & 0.32 & 0.32 \\
39.01 & Y~{\sc ii} & 2.53 & 0.32 & 0.35 & 0.22 & 0.09 & 0.24 & 13 & 10 hfsfit & 2.21 & $$-$$0.05 & 0.37 & 0.43 \\
40.00 & Zr~{\sc i} & 3.48 & 0.90 & 0.93 & 0.06 & 0.14 & 0.15 & 2 & & 2.58 & 0.58 & 0.32 & 0.33 \\
40.01 & Zr~{\sc ii} & 3.45 & 0.87 & 0.90 & 0.25 & 0.06 & 0.26 & 5 & & 2.58 & 0.59 & 0.28 & 0.38 \\
56.01 & Ba~{\sc ii} & 2.10 & $-$0.08 & $-$0.05 & 0.16 & 0.20 & 0.26 & 5 & 5 hfsfit & 2.18 & $$-$$0.61 & 0.53 & 0.55 \\
57.01 & La~{\sc ii} & 1.97 & 0.87 & 0.90 & 0.19 & 0.09 & 0.21 & 11 & 7 hfsfit & 1.10 & 0.49 & 0.38 & 0.42 \\
58.01 & Ce~{\sc ii} & 2.65 & 1.07 & 1.10 & 0.16 & 0.09 & 0.18 & 24 & & 1.58 & 0.73 & 0.34 & 0.38 \\
59.01 & Pr~{\sc ii} & 1.00 & 0.28 & 0.31 & 0.12 & 0.35 & 0.37 & 2 & uncertain & 0.72 & 0.21 & 0.07 & 0.14 \\
60.01 & Nd~{\sc ii} & 2.40 & 0.98 & 1.01 & 0.14 & 0.12 & 0.18 & 24 & & 1.42 & 0.59 & 0.39 & 0.41 \\
62.01 & Sm~{\sc ii} & 2.70 & 1.74 & 1.77 & 0.17 & 0.11 & 0.20 & 24 & & 0.96 & 1.36 & 0.38 & 0.42 \\
63.01 & Eu~{\sc ii} & 2.26 & 1.74 & 1.77 & 0.20 & 0.10 & 0.22 & 8 & 8 hfsfit & 0.52 & 1.45 & 0.29 & 0.35 \\
64.01 & Gd~{\sc ii} & 3.04 & 1.97 & 2.00 & 0.17 & 0.09 & 0.19 & 17 & & 1.07 & 1.65 & 0.32 & 0.36 \\
66.01 & Dy~{\sc ii} & 2.09 & 0.99 & 1.02 & 0.12 & 0.11 & 0.16 & 2 & uncertain & 1.10 & 0.62 & 0.37 & 0.39 \\
68.01 & Er~{\sc ii} & 1.42 & 0.50 & 0.53 & $\cdots$ & 0.09 & 0.09 & 1 & uncertain & 0.92 & 0.18 & 0.32 & 0.32 \\
70.01 & Yb~{\sc ii} & 2.36 & 1.52 & 1.55 & 0.04 & 0.27 & 0.27 & 2 & & 0.84 & 1.21 & 0.31 & 0.31 \\
 \bottomrule
\end{tabular}
\label{tab:abundances}
\end{center}
\end{table*}

For the Li abundance, we treated the Li doublet as a single line because only an upper limit for the equivalent width (0.5 m\AA) could be measured, as noted in the remarks of Table\,\ref{tab:abundances}. The abundances of other doublet or triplet lines in each element were obtained using the program {\sc mpfit} via Gaussian profile fitting. The numbers of corresponding lines are noted in the remarks, such as ``3 triplets", in Table \ref{tab:abundances}. 
Two doublet lines of Na~{\sc i} and one triplet line of Ca~{\sc ii} were also treated as a single lines because of bad profile fitting.
 When hyperfine splitting (hfs) components and relative isotopic fractions of odd nuclei are both available and significant for a given line, the Gaussian profile fitting takes the hfs components and relative isotopic fractions into account. We adopted hfs components and isotopic fractions from Kurucz\footnote{\url{http://kurucz.harvard.edu/linelists/gfhyperall/}}. The numbers of corresponding lines are remarked with ``hfsfit'' in Table \ref{tab:abundances}.
 
The abundances of most elements having $Z>20$ differ by 0.3\,dex between the two different atmosphere models. This difference indicates that the uncertainties on the derived abundances for a given model (i.e. the standard deviations given in column 5 of Table\:\ref{tab:abundances}) are the random uncertainties, only. We conclude that systematic uncertainties also apply but are not accounted for in the {\sc sptool} analysis. We adopt the abundance differences between the two models as the systematic uncertainty, and we combine this in quadrature with the random uncertainty to determine the total uncertainty, given in the final column of Table\:\ref{tab:abundances}. We use this total uncertainty, hereafter.
 
\subsection{Abundance patterns}
\label{sec.pattern}

\begin{figure}
\centering
\includegraphics[width=0.95\columnwidth]{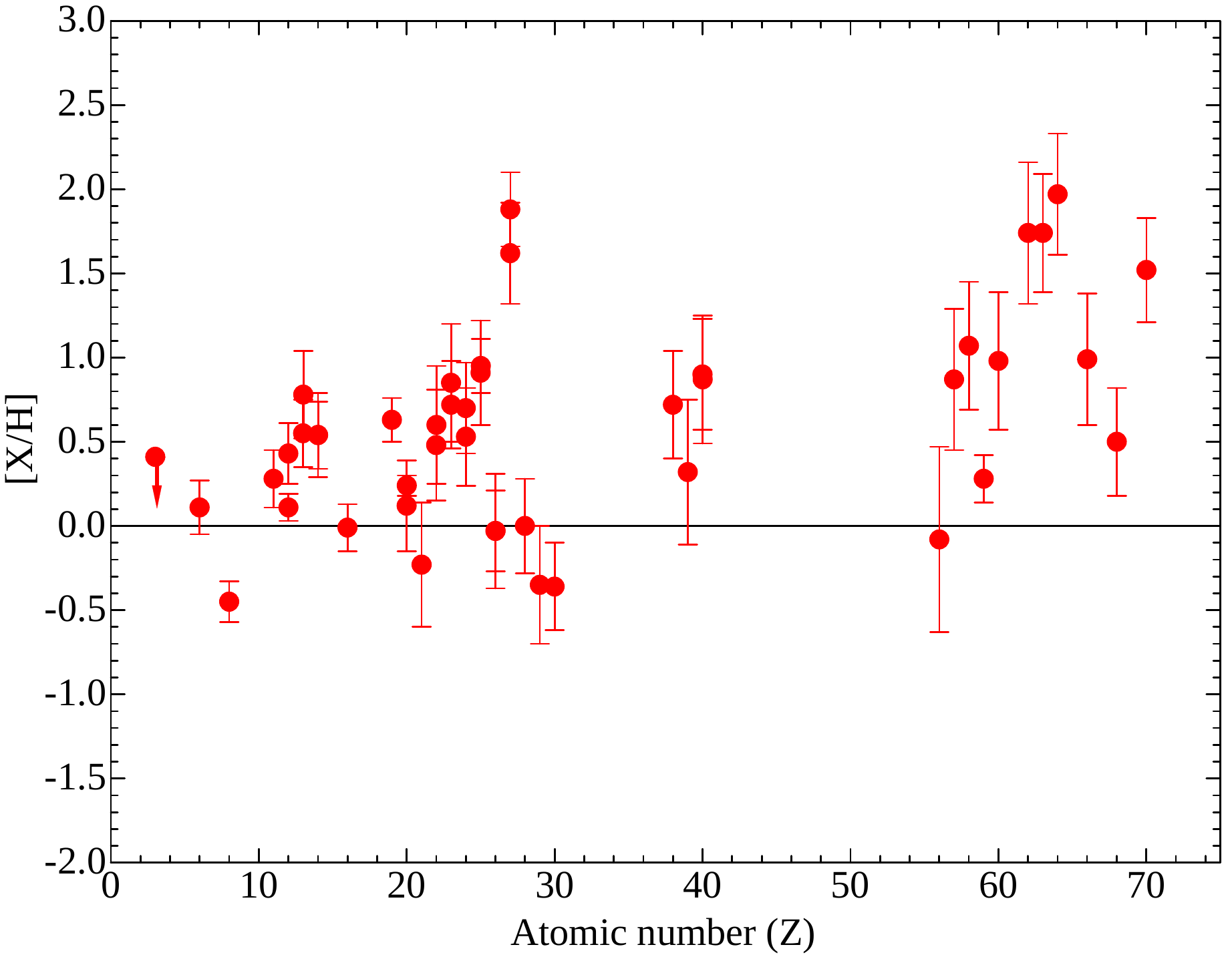}
\caption{[X/H] against atomic number for KIC\,11296437, with one data point for each ion. 
Uncertainties are the total uncertainties from the final column of Table\:\ref{tab:abundances}.}
\label{Fig:sunpattern}
\end{figure}

To illustrate the abundance pattern relative to the Sun, we depict [X/H] against atomic number in Fig.\,\ref{Fig:sunpattern}. 
The light elements from C to Ti, except for O and Sc, show nearly solar abundance or overabundance within 1\,dex. Fe-peak elements V, Cr and Mn ($Z=23, 24, 25$) show overabundances similar to those of light elements, and Co ($Z=27$) is remarkably enhanced by $\sim 1.5$\,dex. Conversely, Fe and Ni are solar in abundance, and Cu and Zn are underabundant. It seems that the odd-even effect is violated for elements from V to Ni, which suggests that atomic diffusion plays an important role in the observed abundance pattern \citep{michaud1970}. Heavy elements vary in their overabundance. 

\begin{figure}
\centering
\includegraphics[width=0.95\columnwidth]{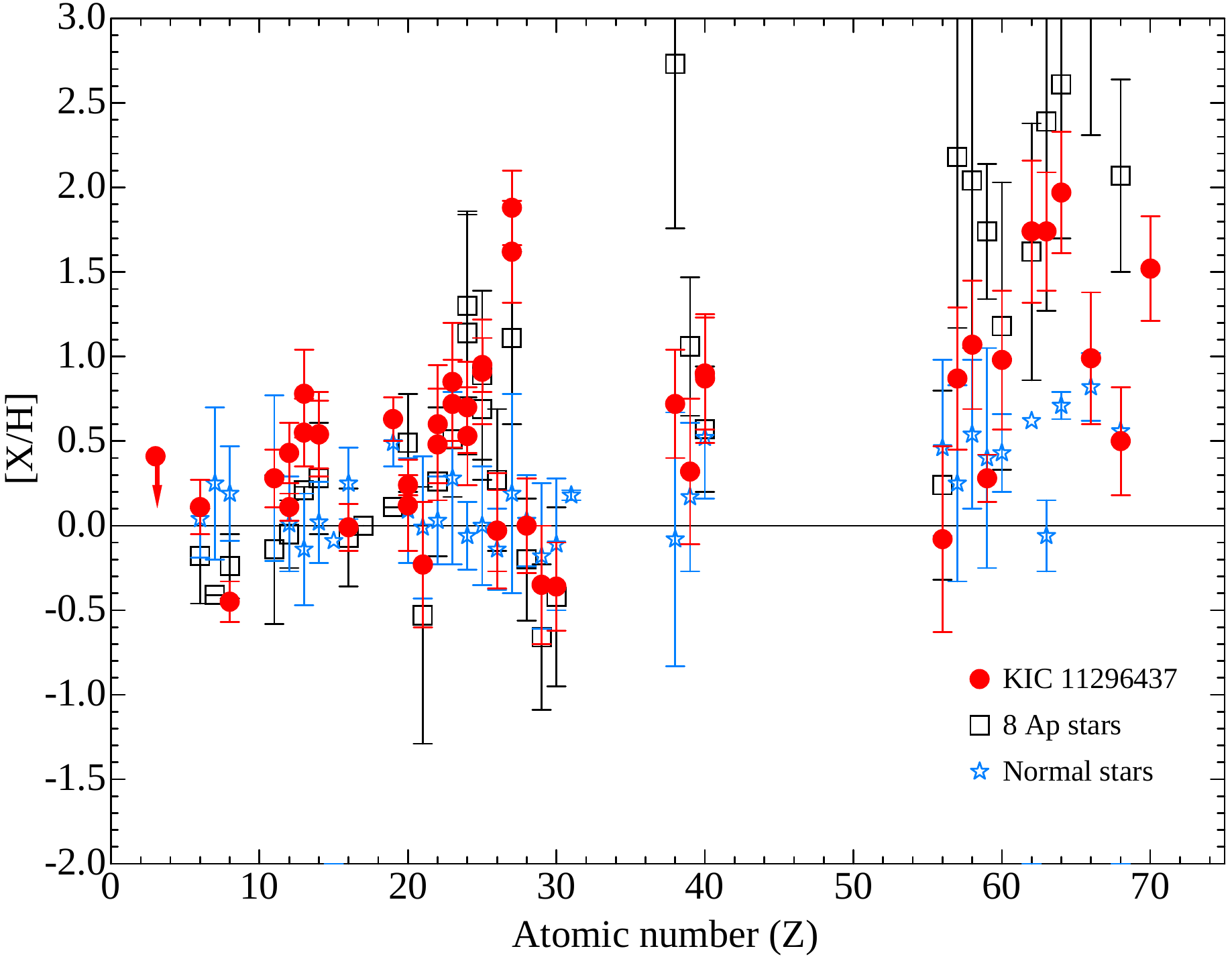}
\caption{Comparison of abundance patterns between KIC\,11296437, 96 normal A/F stars, and the average abundance pattern of eight Ap stars (comprised of seven roAp stars and one $\delta$ Sct star with an Ap chemical signature). KIC\,11296437 has an abundance pattern more consistent with the roAp stars than the normal stars. Uncertainties for KIC\,11296437 are the total uncertainties from the final column of Table\:\ref{tab:abundances}, and for comparison samples are the standard deviations within those samples.}
\label{Fig:appattern}
\end{figure}

\begin{figure}
\centering
\includegraphics[width=0.95\columnwidth]{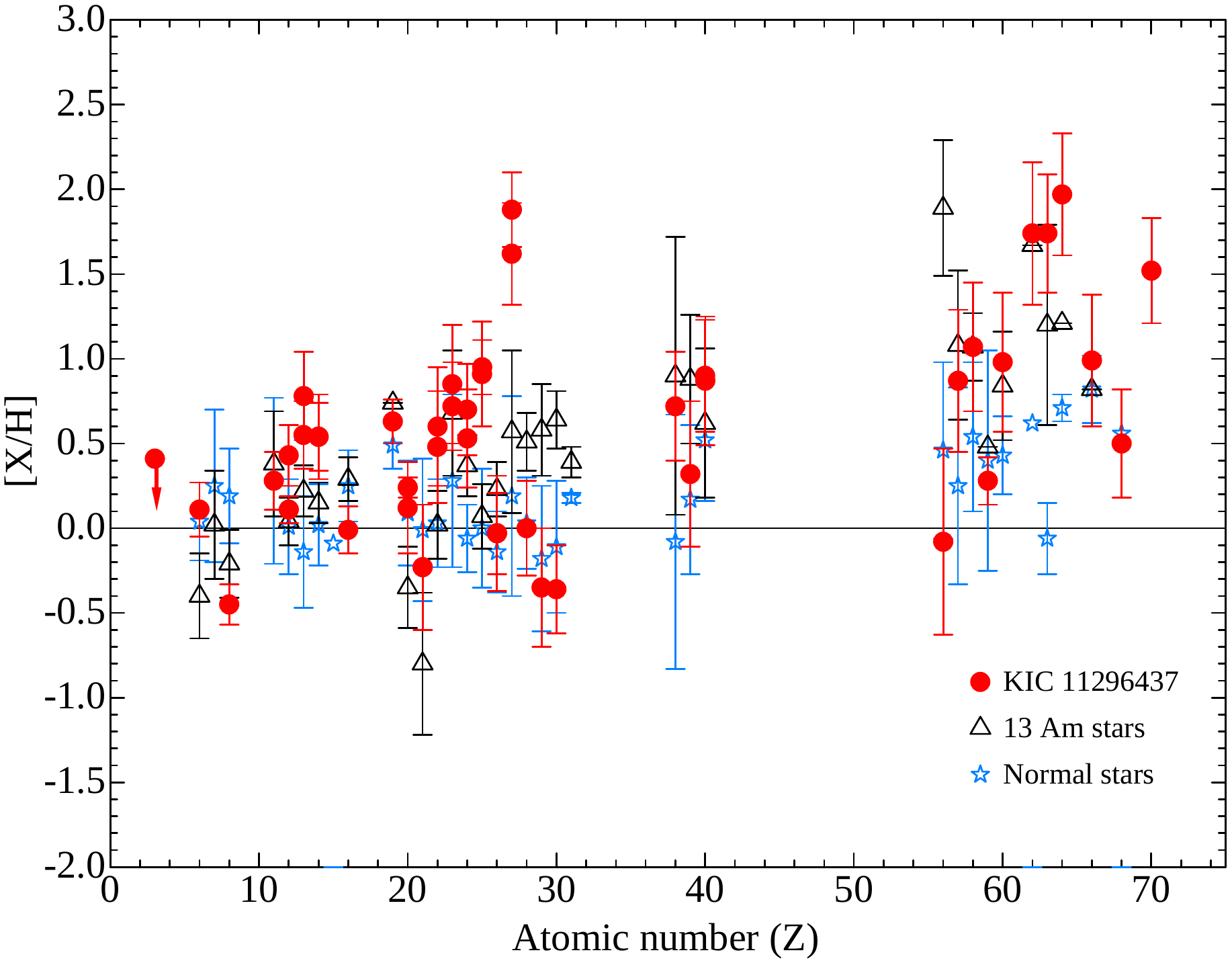}
\caption{Comparison of abundance patterns between our target star, normal A/F, and Am stars. Our target star shows the pattern which is inconsistent with those of Am stars in Fe-peak elements and Ba. Uncertainties for KIC\,11296437 are the total uncertainties from the final column of Table\:\ref{tab:abundances}, and for comparison samples are the standard deviations within those samples.} 
\label{Fig:ampattern}
\end{figure}

It is important to determine whether the diffusion-induced abundance anomalies in KIC\,11296437 are typical of the magnetic Ap stars, or the non-magnetic Am stars. KIC\,11296437 shows both high-overtone roAp p\:modes and low-overtone $\delta$\,Sct p\:modes. Is it the first Am star to show roAp pulsations? Or is it the first roAp star to show $\delta$\,Sct pulsations? To clarify this, we compare the abundance pattern of KIC\,11296437 with those of Ap and Am stars in Fig.\,\ref{Fig:appattern} and Fig.\,\ref{Fig:ampattern}, respectively, We also show normal stars for reference.

\pagebreak

In Fig.\,\ref{Fig:appattern} we compare the abundance pattern of our target star (except for Li)  with those of the average abundances of the 96  normal A--F stars analysed by \citet{niemczuraetal2015}, and the average abundance pattern of eight Ap stars, consisting of one $\delta$\,Sct star with an Ap chemical signature (but no rapid oscillations; HD\,41641 \citealt{escorzaetal2016}), and seven roAp stars: HD\,203932  \citep{1997A&A...319..630G}; 10 Aql,  $\beta$\,CrB and 33 Lib  \citep{2004A&A...423..705R};  KIC\,4768731 \citep{niemczuraetal2015}; HD\,177765 \citep{2012MNRAS.421L..82A};  and $\alpha$\,Cir \citep{2008MNRAS.386.2039B}. The abundance pattern of our target star is more consistent with the roAp stars than with the normal stars. However, certain elements have substantially smaller enhancements than are typical of roAp stars. Sr (Z=38) is strongly overabundant in most Ap stars, but not in KIC\,11296437. The same is true of Pr, Dy and Er (Z=59, 66, and 68).

In Fig.\,\ref{Fig:ampattern} we extend the comparison to the 13 Am stars in \citet{niemczuraetal2015}. While there is agreement among the elements with low and high atomic number, the abundance pattern has an almost opposite behaviour to that of Am stars in some key regards. The Fe-peak elements are a poor match: these elements are all overabundant in Am stars by similar amounts ($\sim$0.5\,dex), whereas in KIC\,11296437 there are substantial over- and under-abundances (1.5 and $-0.5$\,dex) much better matched to the Ap abundance pattern. The abundance of Fe itself is of particular interest: this element is overabundant in Am stars but its abundance in Ap stars is temperature dependent \citep{ryabchikova2005}. In KIC\,11296437, the abundance is not consistent with Am stars, but is consistent with Ap stars of similar temperature. Another conspicuous outlier is Ba. In Am stars, Ba is always strongly overabundant ($\sim$2\,dex; \citealt{niemczuraetal2015,niemczuraetal2017}), whereas in Ap stars it is typically solar \citep{ryabchikovaetal2004}. Its solar abundance in KIC\,11296437 is again more similar to the Ap stars.

In summary, the abundance pattern of KIC\,11296437 is broadly similar to that of the Ap stars, but with only modest overabundances of rare earth elements. It is not consistent with the Am stars. We take this as further evidence of a magnetic field in KIC\,11296437.


\section{KIC\,11296437 is not a binary star}
\label{sec:binary}

One possible explanation for the presence of both roAp and $\delta$\,Sct pulsation in the same light curve is that the target is binary, with one $\delta$\,Sct star and one roAp star. This was the hypothesis that \citet{kurtzetal2008} adopted for the visual binary HD\,218994 (Sec.\,\ref{sec:intro}). However, the high-resolution Subaru spectrum of KIC\,11296437 shows no evidence of being double-lined (SB2), although it is feasible that the $\delta$\,Sct star could be rotating so rapidly ($v_{\rm rot} \sim 300$\,km\,s$^{-1}$) that its rotationally broadened metal lines are impossible to detect at \mbox{S/N $=80$}. In such a case, double-cores might still be expected in some spectral lines, especially those of hydrogen, but none were observed. Furthermore, the measured radial velocity ($-33.0 \pm 0.5$\,km\,s$^{-1}$) agrees with the two in the literature ($-36.3,$\footnote{No uncertainty was given for this particular star, but the average uncertainty given by \citeauthor{frascaetal2016} for their whole sample is 12\,km\,s$^{-1}$.} and $-33.1 \pm 2.0$\,km\,s$^{-1}$; \citealt{frascaetal2016,gaiacollaboration2018a}). We therefore conclude that KIC\,11296437 is neither an SB2 nor SB1 system.

Another diagnostic for binarity is a large renormalised unit weight error (RUWE) in Gaia DR2. Astrometric solutions in DR2 are computed on the basis of single-star motions, so binary stars have excess noise which is encapsulated in the RUWE parameter. Values above 1.4 are considered to be `bad', and objects with RUWE$>$2.0 are likely to be binaries \citep{evans2018,rizzutoetal2018}. KIC\,11296437 has RUWE = 1.121, which argues against this star being a binary.

We applied the phase modulation (PM) method for finding binary stars via the influence of their orbital motion on the stellar pulsations \citep{murphyetal2014}. Using the two $\delta$\,Sct p modes and subdividing the light curve into 10-d subdivisions, we looked for variations in pulsation phase but found none of significance. The null result places an upper limit on any potential companion mass. Assuming an inclination of $90^{\circ}$ and a mass of 1.75\,M$_{\odot}$ for the $\delta$\,Sct star, a companion with a mass > 1.2\,M$_{\odot}$ is ruled out for orbital periods > 25\,d. Unfortunately, the low pulsation amplitudes prevent more stringent constraints, but we note that companions to $\delta$\,Sct stars with even lower pulsation amplitudes have been found with this method before \citep{murphyetal2018}. For an inclination of $60^{\circ}$, which is the median of an isotropic distribution of orbital inclinations, the limit is 1.47\,M$_{\odot}$, whereas the Ap star of lowest known mass, Przybylski's star (HD\,101065), is 1.53\,M$_{\odot}$ \citep{mkrtichianetal2008}. The PM analysis therefore suggests that the $\delta$\,Sct star does not have an roAp companion unless the orbit is at low inclination or at short period. The lack of a periodic signal at long periods other than from the stellar rotation suggests that the target is not an ellipsoidal variable, so a short-period binary is ruled out unless the inclination is low. Importantly, the apparent blind-spot to low inclination is addressed by the Gaia astrometry:
ellipsoidal variability and the PM method are sensitive to motion in the radial direction, which is perpendicular to the astrometric plane. Together, the PM constraint, the light curve, the DR2 RUWE parameter, and the non-detection of line-doubling suggest that KIC\,11296437 is single. 

We also considered that the KIC photometry could be a blend of the light of a $\delta$\,Sct star and an unbound roAp star that just happen to lie close on the sky. Within a search radius of 20$\arcsec$ around KIC\,11296437, there are five other objects in Gaia DR2. The parallaxes show that these other stars are much farther away. In addition, KIC\,11296437 has G = 11.65\,mag; the others have G>17.97\,mag, so contribute very little flux. The other stars are also much redder. KIC\,11296437 has BP$-$RP = 0.5; the others have BP$-$RP>1.0, which makes it unlikely that any of the distant contaminants is an A star capable of $\delta$\,Sct or roAp pulsation. None the less, let us imagine that a substantially reddened background contaminator is the origin of the low-overtone p\:modes. It must be 6 magnitudes fainter, hence invisible in the Subaru spectrum. The observed $\delta$\,Sct oscillations would then have Kp amplitudes around (0.08\,mmag * 10$^{6/2.5}$ * 2) = 40\,mmag intrinsically, where we have used \citeauthor{pogson1856}'s (\citeyear{pogson1856}) magnitude scale and the factor of 2 comes from the fact that if this hypothetical background $\delta$\,Sct star were 6\,mag brighter it would still contribute only half the light. This 40\,mmag amplitude is possible, but unlikely. By inspecting the p-mode amplitudes of the 1988 \textit{Kepler} $\delta$\,Sct stars in \citet{murphyetal2019}, we find that 0.7\:per\:cent have their strongest peak in excess of 40\,mmag. However, most $\delta$\,Sct stars with $T_{\rm eff}$ near 7000\,K also show g modes \citep{uytterhoevenetal2011,bowman&kurtz2018}, and KIC\,11296437 does not. That would be another unusual characteristic of a background star, but no surprise for an Ap star. We therefore argue that the two low-overtone p modes are in the same star as the roAp pulsation.


\section{Asteroseismic models}
\label{sec:seismo}

In this section, we investigate pulsation models that fit the low frequencies observed in KIC\,11296437, and reconcile those with the magnetic field and depleted near-surface helium abundance expected from a slow rotator with weak surface convection.

\subsection{Models to fit low frequencies}
In a first attempt to identify the two pulsation modes, we consider whether they might constitute a rotationally split doublet with the same radial order and degree ($n$ and $\ell$), but different azimuthal order $m$ (specifically $m=\pm1$). From the spot variation in the light curve and from the rotational sidelobes of the roAp dipole mode at 1.41\,mHz (Table\:\ref{tab:freqs}), we measure a rotation frequency of $f_{\rm rot} = 0.1403641 \pm 0.0000004$\,d$^{-1}$ (0.001625\,mHz), which is much smaller than the low-frequency p\:modes, but comparable with the difference between them (0.00236\,mHz). If the low-frequency peaks are a rotationally split doublet, then to a first-order approximation valid for this slow rotator, their separation is $2(1-C_{\ell,n})f_{\rm rot}$, which implies that $C_{\ell,n} = 0.274$. This value is too large for p modes having periods comparable with the fundamental radial mode, and too small for g\:modes. These nonradial modes have \mbox{$C_{\ell,n}\sim 0.04$ to 0.124} (Table~\ref{tab:models}). We conclude that the two observed low frequencies are a radial and a zonal ($m=0$) mode, or two zonal nonradial modes, to explain the lack of additional rotational splittings. The choice of axisymmetry will be discussed in Sec.\,\ref{ssec:non-ad}.

The two observed low frequencies were compared with theoretical frequencies of $\ell=0, 1$ and $2$ modes, for main-sequence evolutionary models with $Z=0.025$, 0.020, 0.015, and 0.010, in which the initial helium abundance in the fully ionized layers was fixed at $Y_0=0.28$.
To mimic the gravitational settling of helium due to slow rotation, we depleted helium above its first ionization zone:
\begin{eqnarray}
Y = 0.01 + (y^+ + y^{++})(Y_0-0.01),
\label{eq:he1}
\end{eqnarray}
where $y^{+}$ and $y^{++}$ are the mass fractions of singly and doubly ionized helium, respectively. Convection in the envelope was suppressed, supposing that a strong magnetic field stabilizes the outer layers, as in the polar model of \citet{balmforthetal2001}. The helium depletion in the surface layers does not affect the stellar evolution; we also found that it hardly affects the driving of low frequency pulsations.

\begin{table}
\begin{center}
\caption{Pulsation frequencies of KIC\,11296437 obtained by \citet{heyetal2019}, used for asteroseismic modelling.}
\label{tab:freqs}
\begin{tabular}{cc}
\toprule
     Frequency &  Amplitude \\
       (mHz)   &     (mmag) \\
\midrule
     1.408152  &  0.020  \\
     1.409777  &  0.352 \\ 
     1.411401  &  0.018 \\
\\
     0.126791  &  0.0222 \\         
     0.129151  &  0.0317  \\ 
\bottomrule
\end{tabular}
\end{center}
\label{tab}
\end{table}

\begin{table}
\caption{Models with suppressed convection and helium depleted to the first He ionization zone having low frequencies similar to those of KIC\,11296437.}
\centering
\begin{tabular}{cccccccc}
\toprule
$M$ & $Z$ & $T_{\rm eff}$ & $\log L/L{\rm _\odot}$ & $\ell$ & $n$ & $\nu$ & $C_{\ell,n}$ \\
M$_{\odot}$ & & K & & & & mHz & \\
\midrule
  1.75 & 0.02 & 6930 & 1.06 &  0   &  0  &  0.1297 \\
          &        &           &        &  2   & -3  & 0.1271  & 0.125 \\
\midrule
  1.80 & 0.02 & 7050 & 1.11 & 1 & -1 & 0.1300 & 0.037 \\
          &        &         &        & 2 & -3 & 0.1268 & 0.108 \\
\midrule
 1.65 & 0.01 & 7211 & 1.13 & 0 &0  &0.1272 & \\
         &       &          &        & 1 & -2 & 0.1293 & 0.107 \\
\bottomrule
\end{tabular}
\label{tab:models}
\end{table}

\begin{figure*}
\centering
\includegraphics[width=0.49\textwidth]{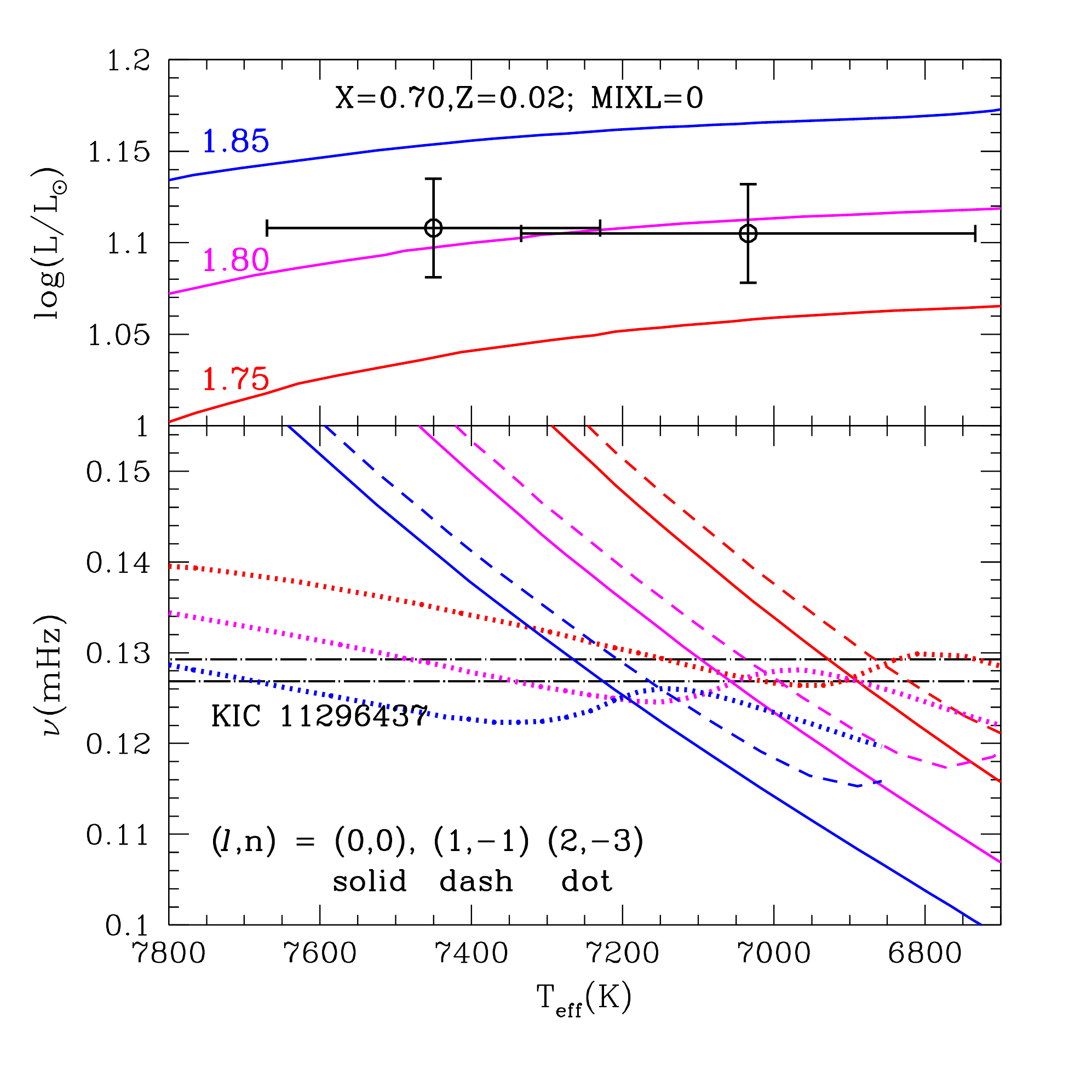}
\includegraphics[width=0.49\textwidth]{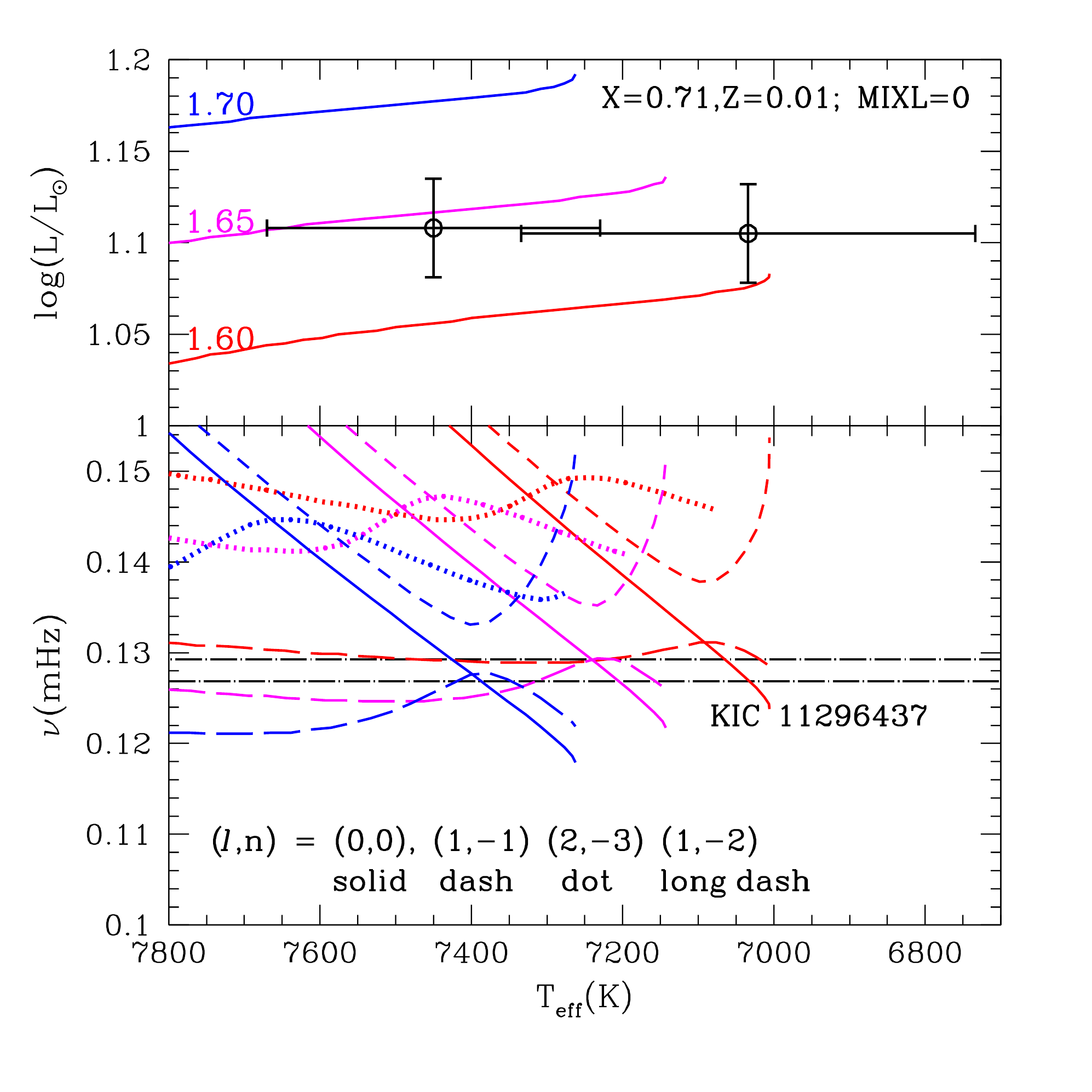}
\caption{{\bf Upper panels:} Evolutionary tracks in the HR diagram for models with three masses. Error bars indicate the locations of KIC\,11296437, described in Sec.\,\ref{sec:atmosphere}. {\bf Left} and {\bf right} panels show models with $(X,Z)= (0.70,0.02)$ and $(0.71, 0.01)$, respectively.
{\bf Lower panels:} Frequencies versus effective temperature for the radial fundamental (solid lines), dipole (dashed lines), and quadrupole modes (dotted lines). Masses are encoded with colour to match the upper panels. Horizontal black dot-dashed lines indicate the two low frequencies observed in KIC\,11296437. The intersection of these horizontal black lines with the coloured lines indicates a frequency match.}
\label{fig:freq}
\end{figure*}

Table\:\ref{tab:models} lists three models that reasonably reproduce the observed frequencies. Only models with $6800<T_{\rm eff}/{\rm K}<7800$ and $1.0<\log L/{\rm L}_{\odot}<1.2$ were searched.
Fig.\,\ref{fig:freq} shows evolutionary tracks and computed frequencies for models with $Z=0.02$ (left panel) and $Z=0.01$ (right panel). The lower panels compare the low frequencies detected in KIC\,11296437 with theoretical frequencies for the radial fundamental and two nonradial modes having frequencies comparable to the observed ones in the selected evolutionary models. For models with $Z=0.02$, the two low frequencies of KIC\,11296437 are consistent with the radial fundamental mode and an $\ell=2$ g\:mode with $n=-3$ at 1.75\,M$_\odot$ ($T_{\rm eff}=6930$\,K), or two g\:modes $(\ell,n)=(1,-1)$ and $(2,-3)$ at $1.80$\,M$_{\odot}$ (7050\,K), where order $n$ is defined as $n_p-n_g$ with $n_p$ and $n_g$ being p-type and g-type radial nodes, respectively (e.g., \citealt{unnoetal1989}). For models with $Z=0.01$, a different g\:mode is preferred.

In these particular models, no overshooting from the convective core is included, but we did investigate the effects of core overshooting, which for a given mass amount to a shift in the evolutionary tracks by 1$\sigma$ in luminosity per 0.1\,$H_p$ of convective core-overshooting. We found that in models of higher overshooting, the fundamental mode has a lower frequency and the frequency difference between it and the other considered modes grows smaller. No good frequency matches were found in models with overshooting, which might be because such overshooting is suppressed by the strong magnetic field in the near-core region of this star. Indeed, \citet{briquetetal2012} and \citet{buysschaertetal2018} found core overshooting was suppressed in the magnetic B stars V2052\,Oph and HD\,43317, respectively.

The depletion of helium and suppression of convection does not affect the pulsation frequencies. For example, a 1.80-M$_\odot$ model with a chemically homogeneous, convective envelope has frequencies of 10.91 and 10.97\,d$^{-1}$ for the modes $(\ell,n)=(0,0)$ and $(2,-3)$, respectively, differing by only $-$0.05 and $+$0.03\,d$^{-1}$ from the model with helium depletion and suppressed convection. The two models have $\Delta T_{\rm eff}<2$\,K.

However, metallicity affects model structure and hence pulsation frequencies considerably. 
For a low-metallicity composition $(X,Z) = (0.71, 0.01)$, a lower mass model (1.65\,M$_\odot$) at higher $T_{\rm eff}$ (7211\,K) has the radial fundamental mode and the $(\ell,n)=(1,-2)$ mode consistent with the observed low frequencies (see Table\:\ref{tab:models} and right panel of Fig.\,\ref{fig:freq}). No reasonable model for the compositions of $Z=0.015$ and $0.025$ could be found.

The $1.41$\,mHz dipole roAp-type frequency of KIC\,11296437 corresponds to radial orders of $n= 28$ and  29 in the 1.75 and 1.80~M$_\odot$ models of Table\:\ref{tab:models}, respectively.  These frequencies are close to the acoustic cut-off frequencies, which is typical for roAp stars \citep{holdsworthetal2018b}. The high-frequency dipole modes are axisymmetric with respect to the pulsation axis, which is aligned with the magnetic axis and thus inclined to the rotation axis. Such axisymmetric pulsations are called oblique pulsations, and appear in the amplitude spectrum with rotational sidelobes separated exactly by the rotation frequency, as first recognized by \citet{kurtz1982}. The high frequency pulsation of KIC\,11296437 is accompanied by a pair of low-amplitude side lobes \citep{heyetal2019}, indicating that the pulsation is an axisymmetric dipole mode aligned with the magnetic axis. The small amplitudes of those side lobes suggest that the obliquity of the magnetic field, $\beta$, and/or the inclination angle between our line-of-sight and the rotation axis, $i$, are small. We showed in Sec.\,\ref{ssec:hires} that the latter is true.

\subsection{Non-adiabatic pulsation analysis}
\label{ssec:non-ad}

Until now, no low-order (low-frequency) pulsations have been detected in roAp stars even though they lie inside the $\delta$\,Sct instability strip. One possible explanation for this is that the strong magnetic field in an roAp star damps low-order pulsations \citep{saio2005}. Pulsational motions in a star couple with magnetic fields most strongly in superficial layers, where the Alfven speed $c_{\rm A}$ is comparable with the sound speed, $c_{\rm s}$.  In deeper layers (where $c_{\rm s} \gg c_{\rm A}$), the acoustic stellar pulsation waves decouple from the magnetic (slow) wave, which is generated by pulsation in upper layers. The slow wave propagates inwards, having progressively shorter spatial wavelengths, and is considered to dissipate before reaching the stellar centre \citep{roberts&soward1983,cunha&gough2000,quitral-manosalvaetal2018}. In other words, in the outer layers of the star, a fraction of the pulsation energy is converted to a magnetic oscillation that propagates inwards and is eventually dissipated in the deep interior, the overall effect of which is to damp stellar pulsation. Here we discuss the strength of magnetic damping on some low-frequency pulsations, as a function of the dipolar field strength at the poles, $B$p.

We assume that pulsation is aligned with the magnetic field axis, and we note that no rotational sidelobes are detected for the two low-frequency pulsations, $\nu_1 = 10.9547$\,d$^{-1}$ and $\nu_2 = 11.1586$\,d$^{-1}$.  If those two frequencies were due to oblique dipole modes, then we would expect the same constraint found by \citet{heyetal2019} of $\tan i \tan \beta =0.11$. That would result in detectable sidelobes, hence we rule out a dipole mode identification for the two low frequencies.

We can look at the geometry of a quadrupole mode quantitatively. From the oblique pulsator model \citep{kurtzetal1990}, the expected amplitude ratios for a quadrupole mode are calculable from $i$ and $\beta$ through:
\begin{equation}
\frac{A^{(2)}_{+2}+A^{(2)}_{-2}}{A^{(2)}_{0}}=\frac{3\sin^{2}\beta \sin^{2}i}{(3\cos^{2}\beta -1)(3\cos^{2}i -1)} 
\end{equation}
\begin{equation}
\frac{A^{(2)}_{+1}+A^{(2)}_{-1}}{A^{(2)}_{0}}=\frac{12\sin\beta \sin i \cos\beta \cos i}{(3\cos^{2}\beta -1)(3\cos^{2}i -1)}.
\end{equation}
From these constraints, and the values of $i < 8^\circ$ and $\beta > 38^\circ$ found in Sec.\,\ref{ssec:hires}, we calculate that the second rotational sidelobes are undetectable for all possible values of $i$ and $\beta$, and that there exists a range of $i$ that can explain non-detection of first rotational sidelobes for quadrupole zonal modes aligned with the magnetic axis.

Thus we conclude that the two low frequencies are plausibly oblique pulsations of a radial and quadrupole mode for which we do not expect to detect rotational sidelobes for this star. We thus proceed with our assumption that pulsation is axisymmetric along the magnetic field axis. Of the models in Table\:\ref{tab:models}, the mode identification favours the 1.75-M$_{\odot}$ model, while the observed luminosity favours the 1.80-M$_{\odot}$ model; in reality, we expect that the star lies somewhere in between but the modelling uncertainties are not small enough to reliably discriminate between a dense grid of models of different metallicity and masses in that range. In what follows, we perform our calculations using the method of \citet{saio2005} and we use the $1.75$-M$_\odot$ model as an indicative model to study and discuss the magnetic damping.

Since the latitudinal dependence of the axisymmetric pulsation cannot be given by a single Legendre function in the presence of a strong magnetic field, the variation due to pulsation, $f(r,\theta,t)$, is expressed as\footnote{We use $i$ as the complex number, rather than the stellar inclination angle, hereafter.}
\begin{eqnarray}
f(r,\theta,t) = e^{{\rm i}\sigma t}\sum_{j=1}^{j=K}f_j(r) N_{\ell_j}P_{\ell_j}(\cos\theta),
\label{eq:exp}
\end{eqnarray}
where $\sigma$ is the complex eigenfrequency ($\sigma=\sigma_r+{\rm i}\sigma_i$), $\ell_j = 2(j-1)$ for even modes and $\ell_j = 2j -1$ for odd modes, $P_{\ell_j}(\cos\theta)$ is the Legendre polynomial with the polar angle $\theta$, and $N_{\ell_j}$ is the normalization factor.  The integer $K$ gives the truncation length of the expansion. Eigenfrequencies and eigenfunctions are obtained by solving $6\times K$ differential equations using the Cowling approximation (see \citealt{saio2005} for details). Convection is suppressed in the models presented in this Section, except in Sec.\,\ref{ssec:other_stars} in which we examine the effect of convection on the excitation of low-order modes.

\begin{figure*}
\centering
\includegraphics[width=0.49\textwidth]{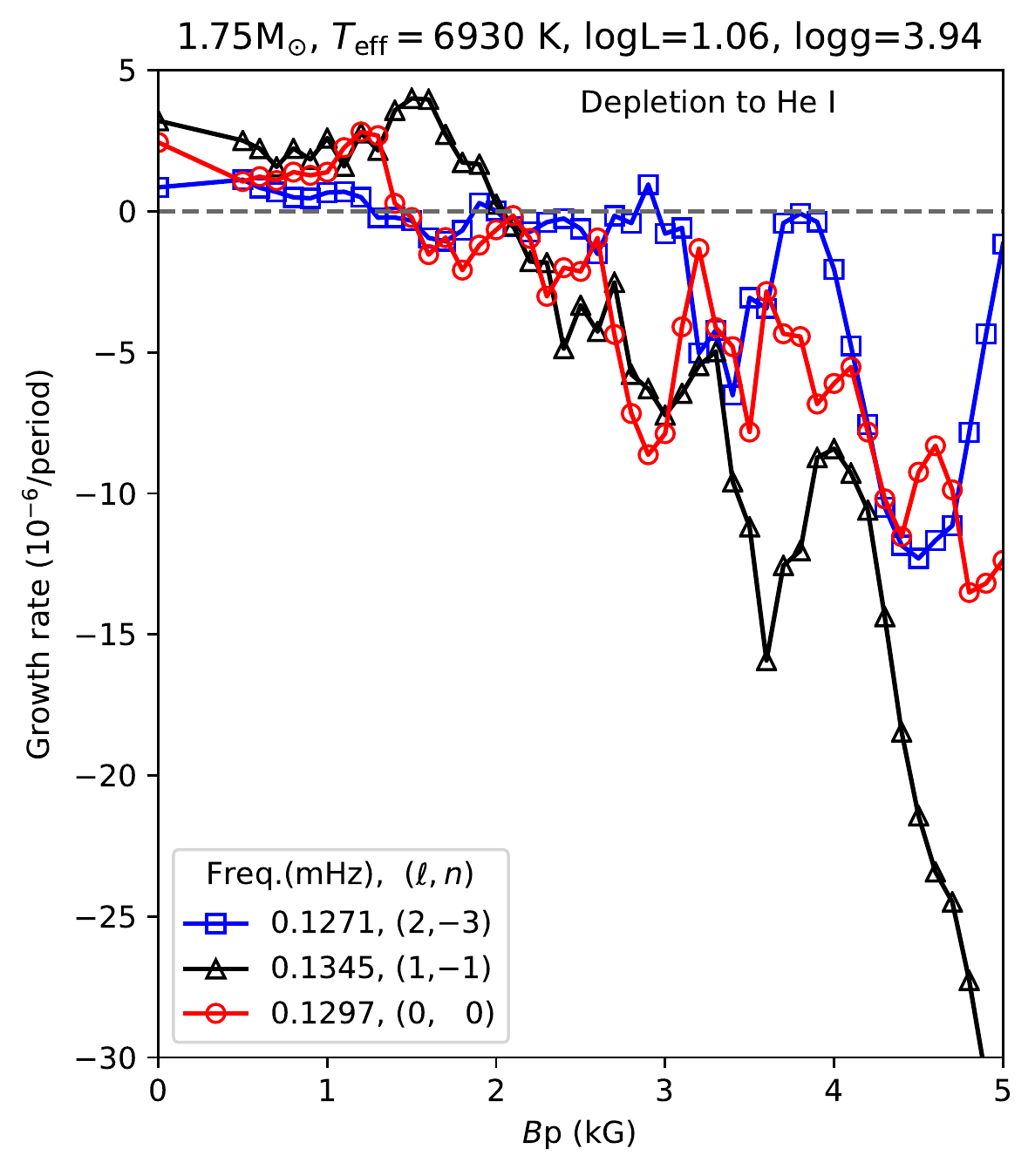}
\includegraphics[width=0.49\textwidth]{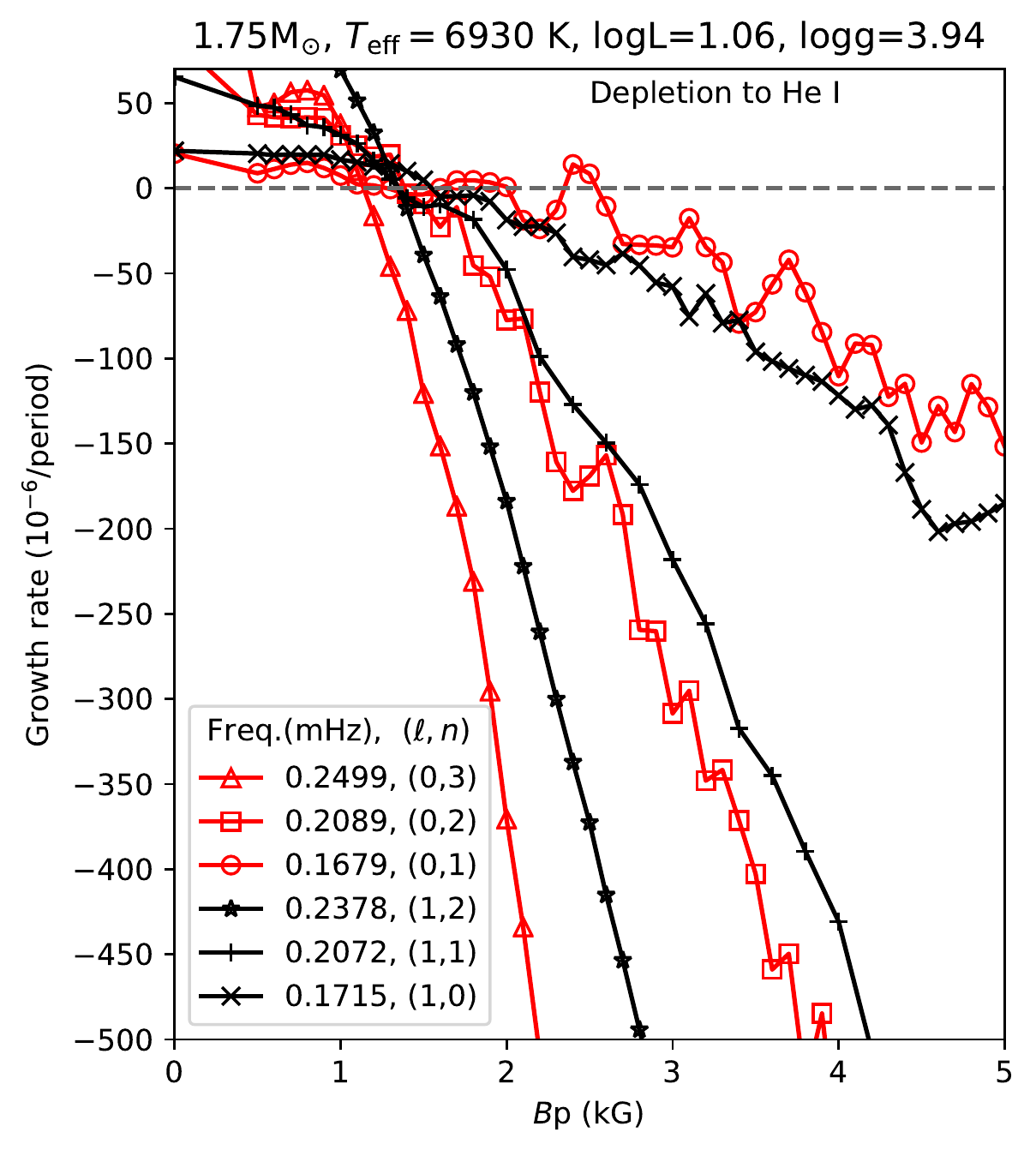}
\caption{Growth rates as a function of magnetic field strength at the poles, $B$p, of pulsation modes in the {\bf $1.75$-M$_\odot$ model at $T_{\rm eff}=6930$\,K} (Table\,\ref{tab:models}), which has He depleted to the first ionization zone.
{\bf Left panel:} Growth rates of the three pulsation modes having frequencies very close to the observed low frequencies of KIC\,11296437. {\bf Right panel}: Growth rates for slightly higher-frequency radial (red) and dipole (black) p\:modes from the same model. Note the different range of growth rates between the two panels.}
\label{fig:growth}
\end{figure*}

For high-order p\:modes (i.e., typical roAp pulsations) \mbox{$K\approx10$--12} is sufficient to obtain reliable eigenfrequencies, where $(|\sigma_i|/\sigma_r)\sim O(10^{-2})$. However, for the low-order p\:modes in KIC\,11296437, $(|\sigma_i|/\sigma_r)\sim O(10^{-6})$,  and the imaginary eigenfrequency components have large scatter as a function of $B$p, even with $K=20$. To reduce the scatter, we averaged the $\sigma_i$ values obtained with $K=12$--25 (12--30 for $\ell=0$ modes of $n = 0$ and 1) at each $B$p. Furthermore, we took three-point running means of the values obtained at every 0.1\,kG. The resulting growth rates ($-2\pi \sigma_i/\sigma_r$) are plotted as a function of $B$p in Fig.\,\ref{fig:growth}. The left panel shows three modes of $(\ell, n)= (0,0), (1,-1)$ and $(2,-3)$, whose frequencies are very close to the low frequencies of KIC\,11296437, while the right panel shows p\:modes with slightly higher frequencies.

For the low frequency modes found in KIC\,11296437, the plotted growth rates still have considerable rapid variations, indicating that the accuracy of the analysis is not sufficient to calculate subtle magnetic effects on the stability of low-order modes. 
With that caveat, Fig.\,\ref{fig:growth} suggests the following:
\begin{enumerate}
\item if $B{\rm p} \lesssim 1.5$\,kG, many low-frequency modes are excited in KIC\,11296437 due to very weak magnetic damping; 
\item if $B{\rm p} \gtrsim 4$\,kG, all low-frequency modes are damped;
\item if $1.5\,{\rm kG} \lesssim B{\rm p} \lesssim 4\,{\rm kG}$, modes with frequencies near the fundamental mode may be excited, while other p\:modes are damped.  
\end{enumerate}
Since the low-frequency pulsations of KIC\,11296437 are identified as $(\ell,n) = (0,0)$ and $(2,-3)$, while p\:modes of slightly higher frequency are not present, our modelling tentatively predicts a polar magnetic field strength between 1.5 and 4\,kG. Given the scatter in the growth rate of $\sim$5$\times10^{-6}$ and accounting for some model misspecification, we infer that this prediction is consistent with the observational estimate of 3.7--4.4\,kG under the assumption of a polar configuration (Sec.\,\ref{ssec:mag}).

For Ap stars more generally, the specific growth rates will of course depend on the stellar properties, but Fig.\,\ref{fig:growth} suggests that p\:modes with the lowest radial orders will experience the least damping. This is because magnetic damping is generally stronger for intermediate order p modes whose kinetic energy mainly lies in the outer layers of the star. So, if the magnetic field is sufficiently weak (say $\lesssim 1\,$kG) and the effective temperature of the star places it in the cooler half of the $\delta$\,Sct instability strip, we expect low-order modes whose periods are not much shorter than the period of the radial fundamental mode. However, we must reiterate that full exploration of the parameter space in $T_{\rm eff}$, $L$ and $B$p has not yet been carried out, and the calculation accuracy needs improvement to facilitate this.
For the high-order and sometimes super-critical p\:modes seen in roAp stars, the excitation mechanism is not yet settled upon \citep{gautschyetal1998,cunha2002,cunhaetal2013}.
However, it is known that the strength of magnetic damping does not increase monotonically with respect to pulsation frequency and $B$p, but varies cyclically \citep{cunha&gough2000,saio&gautschy2004,saio2005}.  This means that there are some ranges of $B$p and oscillation frequency where the magnetic damping is significantly smaller, allowing the excitation of high-order p\:modes in roAp stars.

\subsection{Models with extreme helium depletion}

\begin{figure*}
\centering
\includegraphics[width=0.49\textwidth]{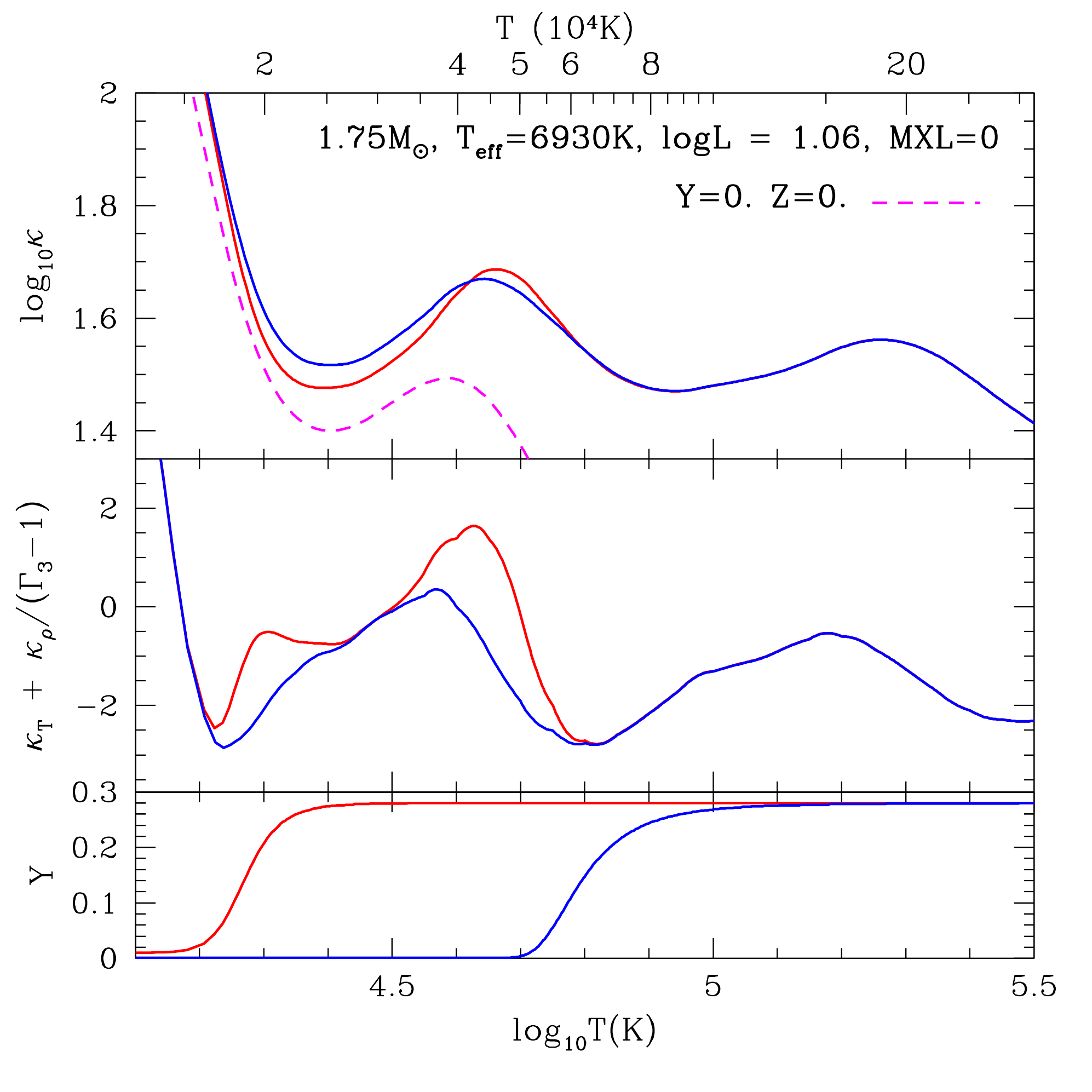}
\includegraphics[width=0.49\textwidth]{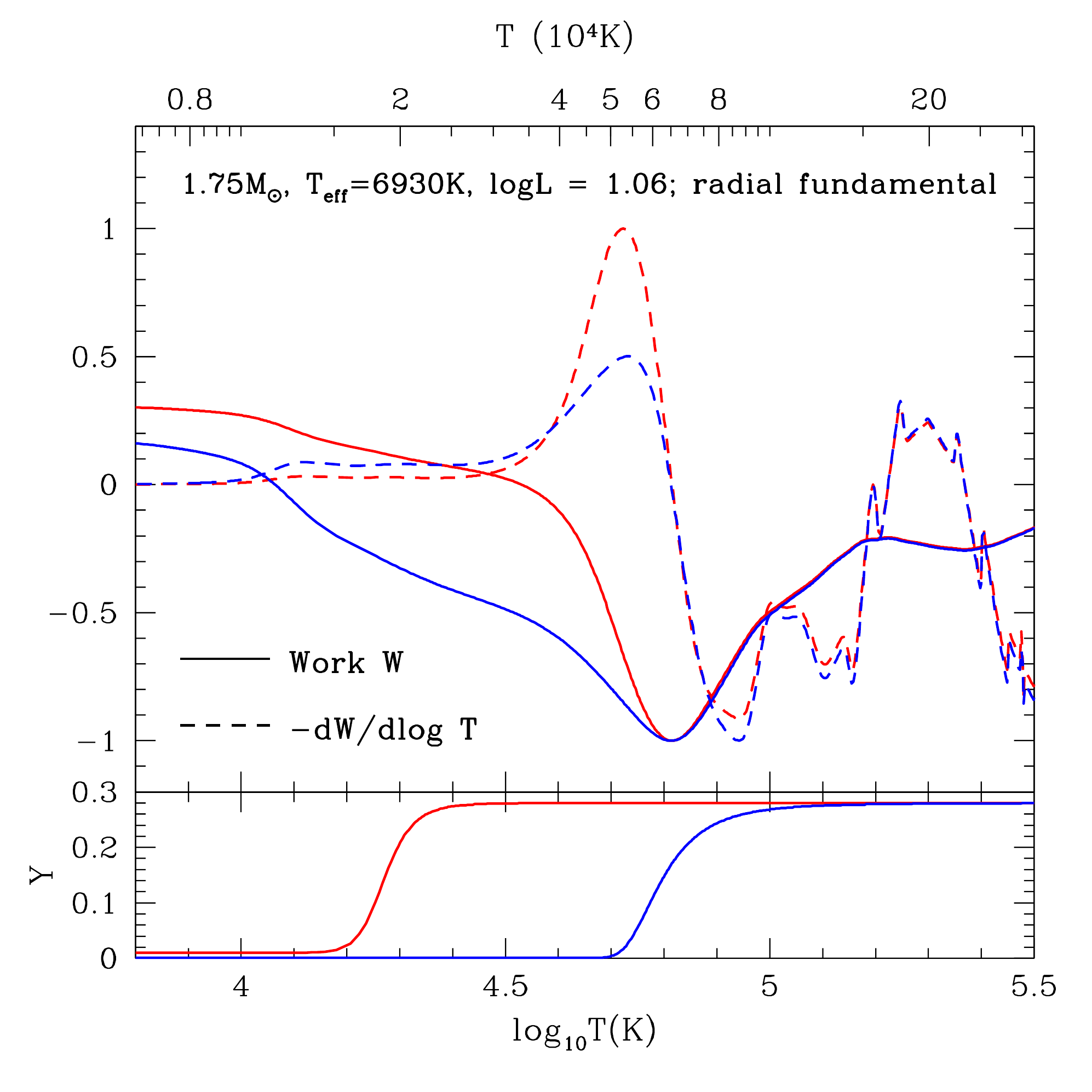}
\caption{{\bf Left, bottom}: helium distribution as a function of temperature: red and blue lines show depletions to the first (eq.\,\ref{eq:he1}) and second (eq.\,\ref{eq:he2}) helium ionization layers, respectively. {\bf Left, middle:} 
$\kappa_{\rm ad}$, which determines the driving (eq.\,\ref{eq:drive}), for the same helium depletion models. {\bf Left, top:} opacity as a function of temperature for the helium depleted models, and the dashed magenta line shows an extreme zero-metallicity model for comparison.
{\bf Right, bottom:} As bottom left. {\bf Right, top:} Accumulated work, $W$, and differential work with respect to $-\log T$, where vertical scales are arbitrary.  Pulsation is driven where $dW/dr > 0$ (or $-dW/d\log T >0$). If $W > 0$ at the surface, the pulsation is excited.}
\label{fig:opc}
\end{figure*}

Pulsational stability is affected by how deeply helium is depleted, although the pulsation periods are hardly affected.
In the models discussed above, helium is depleted only to the first helium ionization zone (eq.\,\ref{eq:he1}). Here we consider models in which helium is depleted to the second helium ionization zone. The local helium abundance is set as
\begin{eqnarray}
Y = 0.001 + (y^{++})^{10}(Y_0-0.001),
\label{eq:he2}
\end{eqnarray}
where the 10th power of $y^{++}$ is arbitrarily adopted to produce near-complete depletion of helium in the second ionization zone. The distributions of helium resulting from eq.\,\ref{eq:he1} and \ref{eq:he2} are compared in the bottom-left panel of Fig.\,\ref{fig:opc}.
Despite the complete depletion of helium in the second ionization zone, there is still an opacity ($\kappa$) bump at almost the same place ($\log T \sim 4.7$) as when helium is depleted only to the first ionization zone (Fig.\,\ref{fig:opc}, top-left). This `edge-bump' \citep{stellingwerf1979} is caused by a discontinuous change in bound-free opacity at the edge of the hydrogen ionization zone. It has gone relatively unnoticed in the literature because the opacity is usually dominated by He II ionization.

The presence of an opacity bump at $\log T\approx 4.6$ in the absence of helium suggests that pulsations of low-order p modes can be excited even in extreme Am stars and non-magnetic or weakly-magnetic Ap stars, in which helium is supposed to be drained from the outer envelope by gravitational settling. Since the effect of this bump in Am/Ap stars has not been discussed in the literature before, we present some details of the $\kappa$-mechanism here.

Pulsational driving via the $\kappa$-mechanism arises if photons from the interior are blocked in the compressed (and hence high-temperature) phase and released in the expanded phase \citep[e.g.,][]{cox1974}. Driving occurs if opacity (in the compressed phase) increases with radius; i.e., for weakly non-adiabatic pulsations, if
\begin{eqnarray}
\frac{d\kappa_{\rm ad}}{dr}>0 \quad \mbox{with} \quad \kappa_{\rm ad}\equiv \kappa_T+\frac{\kappa_\rho}{\Gamma_3-1},
\label{eq:drive}
\end{eqnarray}
where $\kappa_T=(\partial\log\kappa/\partial\log\,T)_\rho$, $\kappa_\rho=(\partial\log\kappa/\partial\log\rho)_T$, and $\Gamma_3$ is the third adiabatic exponent, which arises because the density variation is equal to the temperature variation multiplied by \mbox{$1/(\Gamma_3-1)$} (see \citealt{unnoetal1989} for details).
The middle-left panel of Fig.\,\ref{fig:opc} shows $\kappa_{\rm ad}$ as a function of temperature in the model envelopes in which helium is depleted to the first He ionization zone (Model\,1; red line) and the second ionization zone (Model\,2; blue line).  
The outward increase of $\kappa_{\rm ad}$ around $\log\,T\approx 4.6$ in Model\,2 is slightly smaller than in Model\,1, which indicates the driving at the opacity bump in Model\,2 is weaker than in Model\,1.

Fig.\,\ref{fig:opc} (right) shows pulsational work curves for fundamental radial modes in Model\,1 (red lines) and Model\,2 (blue lines), where driving zones correspond to $-$d$W/$d\,$\log T > 0$ (i.e., d$W$/d$r > 0$). The mode is excited if $W > 0$ at the surface. Comparing driving zones with those from $\kappa_{\rm ad}$ in Fig.\,\ref{fig:opc} (left), we see that driving/damping regions are consistent with the prediction of eq.\,\ref{eq:drive} only for deeper layers having $\log\,T\gtrsim 4.6$, where the pulsation is quasi-adiabatic.  
Other terms in the work integral contribute more in the outer layers where fully non-adiabatic effects play an important role in exciting pulsations, particularly in Model\,2 in which the driving at $\log\,T\sim 4.6$ is weaker. Finally, we note that the opacity bump around $\log\,T\approx 5.3$ (Fig.\,\ref{fig:opc} left) is the well known opacity bump of Fe/Ni ionization, which is not important in driving pulsations in A stars, as can be seen from the small variation of $W$ in Fig.\,\ref{fig:opc} (right).

The excitation of low-order p modes in our models with extreme helium depletion agrees with the earlier analysis by \citet{cunhaetal2004} based on models with helium settling and without winds or magnetic fields \citep{theadoetal2005}. Although no work-curve is shown in their paper, the same `edge-bump' was likely responsible for the excitation.
Interestingly, Cunha (priv. comm.) finds that the loss of driving due to helium depletion affects the fundamental mode most strongly, while we find that the damping effect of the magnetic field affects modes with higher radial order more strongly. Perhaps the combination of these effects is what makes $\delta$\,Sct--roAp hybrids so rare that only one, KIC\,11296437, has been discovered.

\subsection{Application to other stars}
\label{ssec:other_stars}

\begin{figure}
\centering
\includegraphics[width=0.49\textwidth]{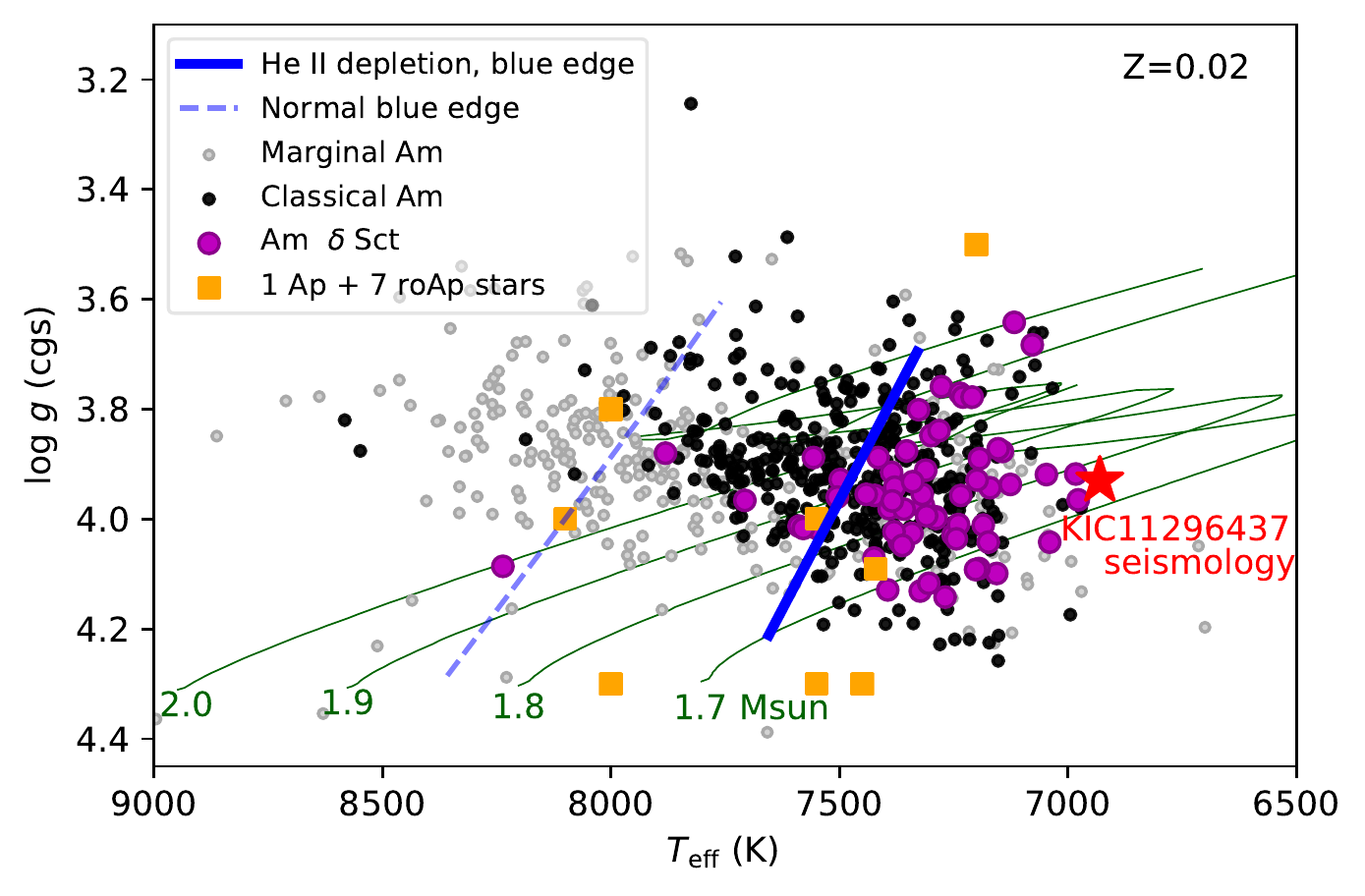}
\caption{The fundamental-mode blue edge for stars depleted of helium to the second He ionization zone. Similar depletion is expected from gravitational settling in Am stars, which are shown according to their degree of peculiarity (from \citealt{smalleyetal2017}). The pulsating Am stars are highlighted, as magenta circles. Evolutionary tracks of helium depleted models at $Z=0.02$ are shown, with their masses written beneath the ZAMS. Orange squares show the positions of the 7 comparison roAp stars in Sec.\,\ref{sec:abundances}, and the post-main-sequence $\delta$\,Sct star with Ap-like abundances. The red star is the seismic result for KIC\,11296437.}
\label{fig:edges}
\end{figure}

For our helium depleted models, we calculated the blue edge of the $\delta$\,Sct instability strip for the fundamental mode, shown in Fig.\,\ref{fig:edges} for models with helium depleted to the second ionization zone. This lies at a cooler temperature than the normal blue edge because of weaker excitation. Similar helium depletion is expected from gravitational settling in Am stars, and our calculated blue edge is in good agreement with observations of pulsating Am stars from \citet{smalleyetal2017}. Those few pulsating Am stars that lie hotter than the blue edge are presumably pulsating in modes of higher radial order,\footnote{Am stars are non-magnetic, so magnetic damping is not relevant, here.} or have incomplete helium depletion, perhaps because they are rotating less slowly than other class members. The gap between the observed population of Am stars and the ZAMS presumably arises because of the time taken for rotational braking and the development of peculiarities via atomic diffusion. It is perhaps unsurprising that the hottest Am pulsator is also relatively young, meaning helium depletion is still incomplete, allowing stronger mode excitation and detectable pulsation.

We examined the effect of convection on the excitation of low-order radial modes by calculating envelope models with no helium, in which we used a local mixing-length theory \citep{henyeyetal1965} with a mixing-length of 1.8 times the pressure scale height. Convection--pulsation coupling was included by adopting the time-dependent convection theory of \citet{grigahceneetal2005}. From the calculations, we found that the blue edge of the fundamental radial mode is shifted by around $-100$ K in $T_{\rm eff}$ while the blue-edge of the fifth overtone mode is located slightly hotter than the solid blue line in Fig.\,\ref{fig:edges}, together indicating that the impact of convection is minor. 
Our blue edge lies near that of \citet{antocietal2019}, who employed a different convection theory and used models with a smaller helium depletion. However, \citet{antocietal2019} found turbulent pressure to be the mechanism driving p\:modes in Am stars, whereas we find that the `edge-bump' is able to drive these modes without turbulent pressure and with more extreme helium depletion. A detailed comparison over a broad parameter space is required, but is beyond the scope of this paper.

Our modelling predicts that more Ap stars (including roAp stars) will be found to pulsate in low-order p\:modes if they are located among the pulsating Am stars in Fig.\,\ref{fig:edges}, and if they have polar field strengths $B{\rm p} \lesssim 4$\,kG. However, we reiterate that the strength of driving and sensitivity to the polar field strength will be a function of other model parameters, such as temperature and mass, and this dependence has not been explored. While the seven roAp stars from Sec.\,\ref{sec:abundances} have appropriate field strengths, they are not well-matched to the locus of pulsating Am stars (Fig.\,\ref{fig:edges}). Nonetheless, we searched for low-overtone modes among the three that have space photometry (KIC\,4768731, HD\,128898, HD\,203932), but no such modes were found. The $\delta$\,Sct star with Ap-like abundances, HD\,41641 \citep{escorzaetal2016}, is quite far above the terminal-age main-sequence (it is the Ap star with lowest $\log g$ in Fig.\,\ref{fig:edges}), beyond the range of our instability calculations. The curation of a sample of Ap stars near the locus of the pulsating Am stars would be highly valuable, and a search for low-overtone p\:modes among them would help to refine our understanding of pulsational driving in stars with near-surface depletions of helium.

\begin{figure}
\centering
\includegraphics[width=0.49\textwidth]{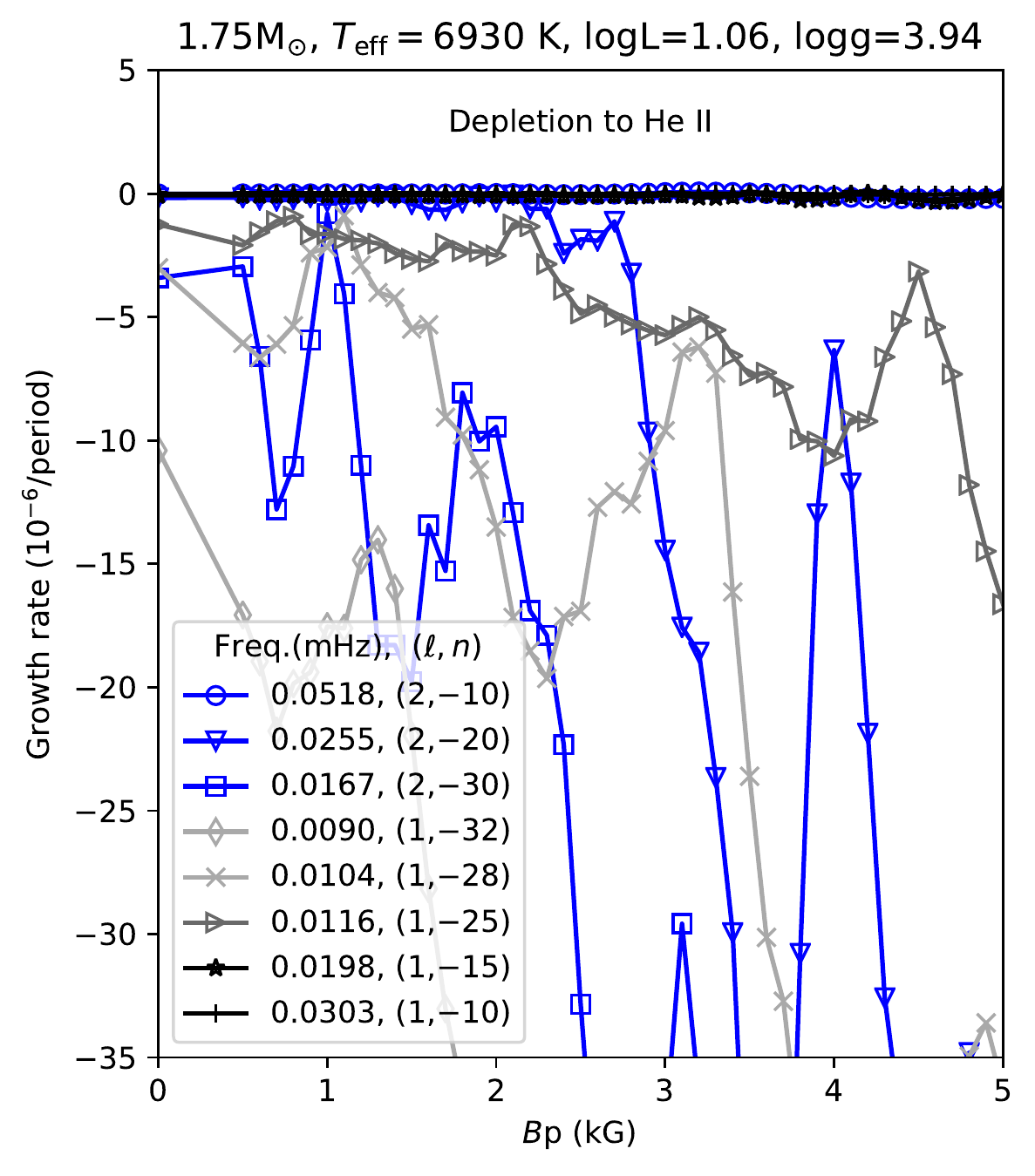}
\caption{Growth rates for a range of $\ell=1$ and 2 g\:modes as a function of polar field strength, $B{\rm p}$. The ($\ell,n$) = ($1,-10$) and ($1,-15$) modes lie very close to zero, along with the ($2,-10$) mode. Radial orders of $-40$ to $-70$ are not shown because the damping is extremely strong, even for weak fields.}
\label{fig:gmodes}
\end{figure}

Using the same extreme helium depletion models, we considered the driving of g\:modes in magnetic Ap stars as well. Observations of 611 \textit{Kepler} $\gamma$\,Dor stars have shown that their radial orders are typically between $-$20 and $-$70 (given by the FWHM for the $\ell=1$ g-mode distribution in \citealt{glietal2020a}), while the g\:modes with frequencies similar to p\:modes have $-1 > n > -10$. In Fig.\,\ref{fig:gmodes} we show growth rates for g\:modes in both regimes for our seismic model of KIC\,11296437, as a function of field strength $B{\rm p}$. We find that low-order g\:modes are not damped by magnetic fields, which is consistent with our mode identification for KIC\,11296437. This is probably because the energy of these modes lies mainly in the deep interior, while magnetic interaction occurs only in the outermost layers of our model. It may be the case that rare examples of $\delta$\,Sct pulsation in Ap stars are examples of low-order g-mode pulsation, rather than p\:modes. Conversely, the high-order g\:modes commonly observed in $\gamma$\,Dor stars are heavily damped by fields stronger than \mbox{1--4\,kG}, with the damping being stronger for higher radial orders. The strong damping can perhaps be attributed to the dominant horizontal motions of such g\:modes, which would strongly disturb magnetic field lines. Furthermore, if such g\:modes are driven by the blocking of convective flux \citep{guziketal2000b}, then the suppression of convection by magnetic fields would lead to weak driving in addition to the strong damping. We therefore explain the observation that no magnetic Ap stars have been observed as $\gamma$\,Dor pulsators.

Ours is not the only explanation for the absence of g\:modes in Ap stars. Our magnetic damping calculations for g\:modes assumed a global dipole magnetic field of a few kG that couples with g\:modes in the envelope. If one assumes that field intensity scales as \mbox{$B{\rm p} \propto r^{-3}$,} the field strength outside the convective core where g\:modes have larger amplitudes ought to exceed $\sim$100\,kG, and might reach 1000\,kG if the field is dipolar all the way to the core. \citet{cantielloetal2016} calculated that fields of $\sim$100\,kG are sufficient to damp g\:modes outside the convective core, hence g\:modes may be damped both at the surface and near the core. Whether the near-core or the surface damping dominates is not clear; a search for g\:modes affected by slightly sub-critical near-core fields \citep{vanbeecketal2020} would be a good test of this.


\section{Conclusions}

Our goal in this work was to determine if KIC\,11296437 is a magnetic Ap star and to measure its magnetic field. From an equivalent width analysis, we measured $\langle H \rangle = 2.8\pm0.5$\,kG, though no Zeeman splitting was observed. We have measured elemental abundances for KIC\,11296437, which we showed are consistent with its classification as an Ap star. The $\alpha^2$~CVn variability seen in Fig.\,\ref{fig:intro}, along with the abundance analysis and magnetic field strength measurement, show that the star is an Ap star, hence it is a high-overtone roAp star and simultaneously a low-overtone $\delta$\,Sct star. We have made a strong case that the two classes of variability are not manifested in different stars of a binary system, by ruling out that the star is a binary over a wide parameter space.

New model calculations indicate that magnetic field strengths less than 1\,kG do not suppress low-overtone p\:modes, and, within the model uncertainties, the fundamental radial mode can be excited even in stars with field strengths up to $\sim$4\,kG. Other roAp stars with low field strengths should also be found with $\delta$\,Sct pulsations, similar to KIC\,11296437. The same calculations show that low-order g\:modes with frequencies similar to the fundamental mode are not suppressed, while the high-order g\:modes typical of $\gamma$\,Dor stars are strongly suppressed, explaining the absence of observations of $\gamma$\,Dor pulsation in Ap stars.

We show that in stars with extreme helium depletion down to the second helium ionization zone, low-order p\:modes can still be driven by a bump in Rosseland mean opacity caused by the H-ionization edge. Gravitational settling of helium is therefore not a barrier to pulsation in Ap stars, regardless of magnetic field strength or rotation rate. We considered the implications for all slowly-rotating A stars that experience gravitational settling, and found that the distribution of Am stars with $\delta$\,Sct pulsation is well explained by the blue edge of the instability strip calculated with our extreme helium depletion models.

\section*{Acknowledgements}

The authors thank the anonymous referee for their careful reading of the manuscript, and thank Jim Fuller, Margarida Cunha and Coralie Neiner for their comments. SJM was supported by the Australian Research Council through DECRA DE180101104. DWK was supported by the Hunstead Gift for Astrophysics at the University of Sydney.
This work has made use of data from the European Space Agency (ESA) mission {\it Gaia} (\url{https://www.cosmos.esa.int/gaia}), processed by the {\it Gaia} Data Processing and Analysis Consortium (DPAC, \url{https://www.cosmos.esa.int/web/gaia/dpac/consortium}). Funding for the DPAC has been provided by national institutions, in particular the institutions participating in the {\it Gaia} Multilateral Agreement. This work also made use of ARI's Gaia Services at \url{http://gaia.ari.uni-heidelberg.de/} for RUWE values.

\section*{Data Availability}
The \textit{Kepler} lightcurve of KIC\,11296437 is publicly available from MAST.\footnote{\url{https://mast.stsci.edu/}} The HDS@Subaru spectrum is obtainable via SMOKA.\footnote{\url{https://smoka.nao.ac.jp/}}

\bibliographystyle{mnras}
\interlinepenalty=10000
\bibliography{11296437.bib}

\bsp	
\label{lastpage}
\end{document}